\newif\ifcheckpagelimits
\checkpagelimitstrue
\checkpagelimitsfalse

\ifcheckpagelimits
 \documentclass[nofootinbib,pra,aps,twocolumn,showpacs,showkeys,%
 amsmath,amssymb,superscriptaddress,final,reprint,floatfix,longbibliography]{revtex4-1}
 \newcommand{\todo}[1]{}
\else
 \documentclass[pra,aps,twocolumn,showpacs,showkeys,%
 amsmath,amssymb,superscriptaddress,final,reprint,floatfix,longbibliography]{revtex4-1}
 \newcommand{\todo}[1]{{\pdfmargincomment[icon=Note,color=pink]{#1}}}
\fi

\usepackage{lineno}
  \usepackage{mathptmx}
\usepackage{subfigure}
\usepackage{dcolumn}
\usepackage{amsmath,amssymb}
\usepackage{bm}
\usepackage{color}
\usepackage{overpic}
\usepackage{latexsym}
\usepackage{epstopdf}
\usepackage{color}
\usepackage[english]{babel}
\usepackage{latexsym}
\usepackage{stmaryrd}
\usepackage{relsize}
\usepackage{graphicx}
\usepackage{braket}

\definecolor{mygrey}{gray}{0.35}
\definecolor{myblue}{rgb}{0.2,0.2,0.8}
\definecolor{myzard}{cmyk}{0,0,0.05,0}
\definecolor{mywhite}{rgb}{1,1,1}
\definecolor{myred}{rgb}{1,0.,0.3}

\usepackage[colorlinks=true,citecolor=myblue,linkcolor=myred]{hyperref}

 \def\ee{\mathord{\rm e}}
 
 \def\ii{\mathord{\rm i}}

\def\half{\textstyle\frac{1}{2}}

\def\fourth{\textstyle\frac{1}{4}}

\renewcommand{\ii}{{\rm i}}
\renewcommand{\ee}{{\rm e}}
\def\beq{\begin{equation}}
\def\eeq{\end{equation}}

\usepackage{amsmath}
\usepackage{graphicx}
\usepackage{calc}
\newlength{\depthofsumsign}
\newlength{\totalheightofsumsign}
\newlength{\heightanddepthofargument}

\newcommand{\nsum}[1][1.4]{
    \mathop{%
        \raisebox
            {-#1\depthofsumsign+1\depthofsumsign}
            {\scalebox
                {#1}
                {$\displaystyle\sum$}%
            }
    }
}

\begin{document}

\title{Micromotion-enabled  improvement of quantum logic gates with trapped ions}

\author{Alejandro Bermudez}
\email[Email to: ]{bermudez.carballo@gmail.com}
\affiliation{Department of Physics, College of Science, Swansea University, Singleton Park, Swansea SA2 8PP, United Kingdom}
\affiliation{Instituto de F\'{i}sica Fundamental, IFF-CSIC, Madrid E-28006, Spain}

\author{Philipp Schindler}
\affiliation{Institute for Experimental Physics, University of Innsbruck, 6020 Innsbruck, Austria}

\author{Thomas Monz}
\affiliation{Institute for Experimental Physics, University of Innsbruck, 6020 Innsbruck, Austria}

\author{Rainer Blatt}
\affiliation{Institute for Experimental Physics, University of Innsbruck, 6020 Innsbruck, Austria}
\affiliation{Institute for Quantum Optics and Quantum Information of the Austrian Academy
of Sciences, A-6020 Innsbruck, Austria}

\author{Markus M\"uller}
\affiliation{Department of Physics, College of Science, Swansea University, Singleton Park, Swansea SA2 8PP, United Kingdom}

\begin{abstract}
The micromotion of ion crystals confined in Paul traps is usually considered  an inconvenient nuisance, and  is thus typically minimised in high-precision experiments such as  high-fidelity quantum  gates for quantum information processing. In this work, we introduce a particular scheme where this behavior can be reversed, making  micromotion   beneficial for quantum information processing.  We show that using laser-driven micromotion sidebands, it is possible to engineer state-dependent dipole forces with a reduced effect of off-resonant couplings to the carrier transition. This  allows one, in a certain parameter regime,  to devise entangling gate schemes based on geometric phase gates with both a higher speed and a lower error, which is  attractive in light of current efforts towards fault-tolerant  quantum information processing. We discuss the prospects of reaching the  parameters  required to observe this micromotion-enabled improvement  in experiments with current and future trap designs. 

\end{abstract}

\ifcheckpagelimits\else
\maketitle
\fi
\setcounter{tocdepth}{2}
\begingroup
\hypersetup{linkcolor=black}
\tableofcontents
\endgroup

 \ifcheckpagelimits
\else
\section{{\bf Introduction}}

The possibility of harnessing the distinctive behavior of quantum-mechanical systems  to process information in new ways has raised the interest of researchers for more than three decades now. This has  given rise to the multi-disciplinary field of {\it quantum information processing} (QIP)~\cite{book_qi}, which could, for instance, have an impact on  current cryptographic protocols~\cite{shor_alg}, or revolutionise our approach to solve long-standing problems in quantum many-body physics~\cite{qs_feynman}. Motivated by such remarkable applications, QIP has now turned into a mature field where experimentalists are using different  technologies~\cite{qc_review}  to face the  challenge  of building registers of ever-increasing sizes,  while trying to preserve and manipulate their quantum features for ever-longer  periods of time.

Among these so-called {\it quantum technologies}, crystals of trapped and laser-cooled  atomic ions~\cite{nist_review,rmp_leibfried,innsbruck_review} have played a leading role in the progress of QIP. Pioneering proposals to build a quantum-information processor based on trapped ions~\cite{cz_gates}, and successfully implemented in the laboratory~\cite{cz_gates_exp}, have opened an active avenue of research with the ultimate goal of building a large-scale trapped-ion quantum computer~\cite{ions_qc_scale}. As emphasised  early in the literature~\cite{landauer}, success in such an enterprise would require {\it (i)} a careful assessment of the possible imperfections of the quantum processor, which lead to  errors in the computation, and {\it (ii)} a thorough study of the unavoidable coupling to an external environment, which degrades the quantum coherence responsible for the advantages of QIP. The former  yields errors that can accumulate quickly  since the information is not stored in classical binary variables~\cite{landauer}, whereas the latter yields an exponential decrease of  quantum coherence with    the size of the register~\cite{unruh_decoherence}.

Despite such a daunting perspective, the subsequent development of quantum error correction showed that these difficulties can be overcome if {\it (i)} one encodes the information  redundantly in an enlarged quantum mechanical system instead of using a bare quantum register, and {\it (ii)} errors are detected and corrected during the storage and processing of encoded states~\cite{qec}. Increasing levels of protection against noise can be achieved e.g.~by concatenating elementary quantum error correcting codes, or by storing logical states in global, topological properties of larger quantum many-body  systems~\cite{topological_codes}. It has been shown that {\it fault-tolerant} QIP is possible provided that errors, either due to imperfections of gates or to environmental decoherence, occur below certain critical rates.  The particular threshold values depend on the details of the implementation, the noise model, and the  chosen encoding. For circuit noise models, a common estimate for concatenated codes is around $10^{-4}$~\cite{threshold_numbers}, whereas topological codes typically offer higher error thresholds up to about $10^{-2}$~\cite{topological_codes}. Essentially, once below the threshold, quantum error correction  allows for slowing down the occurrence of  errors at the  level of the  logical qubits,  such that  longer  computations can tolerate   noise on the  physical qubits  at a  much higher rate. Trapped ions have already demonstrated remarkable  progress in experimental demonstrations of quantum error correction~\cite{qec_ions,color_code_ions,qe_detection}.

In order to meet such a threshold, one must optimise the hardware (i.e. quantum technology) and the software (i.e. schemes to manipulate the quantum information), which can be understood as a built-in {\it error suppression}.  At the software level, one can mitigate decoherence by encoding the information in a section of the Hilbert space that is more robust to the typical environmental noise, as occurs for decoherence-free subspaces~\cite{dfs,dfs_ions}, and for the so-called clock-state qubits~\cite{clock_states}. Regarding imperfections of gates, pulsed~\cite{pulsed_dd, pulsed_dd_ions} and continuous~\cite{continuous_dd,continuous_dd_ions} dynamical decoupling have also been implemented in ion traps. Another possible source of error arises  in certain quantum technologies that exploit additional auxiliary (quasi)particles to mediate an entangling gate between distant qubits, since quantum/classical fluctuations affecting these (quasi)particles   can introduce  errors in the computation. Such is the situation with trapped ions, 
where phonons serve as a quantum bus to generate entanglement, and thermal fluctuations lead to significant errors  when the ion crystals are not laser cooled to the groundstate~\cite{cz_gates}. In this respect, the development of gate schemes  that minimise such thermal sensitivity has been of paramount importance to the field. These schemes typically use a {\it state-dependent dipole force} in the resolved-sideband regime, which forces the ions along a closed trajectory in phase space  depending on the  state of the qubits, either in the  $\sigma^z$~\cite{gphgates,didi_gate} or  $\sigma^\phi$~\cite{ms_gates,ms_gates_nist} eigenstate basis, where $\sigma^\phi = -\cos(\phi) \sigma^y - \sin(\phi) \sigma^x$. This effectively leads to state-dependent multi-qubit geometric phases that can be exploited to generate entanglement,  which underlies the remarkably low errors that have been achieved in experiments so far~\cite{innsbruck_MS,oxford_MS,nist_MS}, with infidelities reaching values below $10^{-3}$. 

An increase of  gate speed yields another clear route for further error suppression, as the  environmental decoherence affecting the qubits, or other external sources of noise affecting the phonon bus, would have a smaller impact during a shorter computation.  
Schemes for ultra-fast entangling gates based on concatenated resonant state-dependent kicks have been studied in detail~\cite{fast_gates_kicks}, which abandon the resolved-sideband regime  to avoid the associated limitations on the gate speed. These schemes 
 give a clear advantage provided that high laser repetition rates~\cite{error_repetition_rate}, and small laser intensity fluctuations~\cite{kicked_gates_error}, can be achieved in the laboratory. Pulse splitting techniques have been implemented in order to increase the number of pulses incident on the ion~\cite{kicks_interferometry},  increasing thus the    repetition rate towards a regime where  ultra-fast gates are expected to have  small errors~\cite{error_repetition_rate}. To overcome the stringent conditions on the laser intensity stability~\cite{kicked_gates_error}, dynamical decoupling approaches may have to be applied in order to minimise the error of each resonant state-dependent kick~\cite{kicks_refocusing}. 
 
 In order to avoid these technical difficulties, but still get an increase on  gate speed with respect to previous realisations~\cite{ms_gates_nist,didi_gate}, schemes  based on state-dependent $\sigma^z$-forces  with an increased  laser intensity have also been studied~\cite{pulsed_sz_gates}, which take into account the leading-order corrections as one abandons the resolved-sideband regime. In this case, such corrections correspond to a time-dependent ac-Stark shift, which is usually  neglected in the resolved-sideband limit~\cite{didi_gate}, but starts contributing as one increases the laser power, and thus the gate speed~\cite{pulsed_sz_gates}. The particular form of the $\sigma^z$-force allows one to take into account this term easily,  finding robust pulse sequences for  faster quantum gates~\cite{pulsed_sz_gates}. Unfortunately, the state-dependent laser forces of this scheme {\it (i)}  cannot be implemented with clock-state hyperfine qubits~\cite{clock_z_forces}, and {\it (ii)} have some limitations for optical qubits in comparison to the entangling gates  generated by  $\sigma^\phi$-forces~\cite{ent_gates_opt_qubits}. It would be thus desirable to consider schemes to speed up entangling gates based on $\sigma^{\phi}$-forces valid for both  hyperfine and optical qubits. Unfortunately, the leading-order corrections to the resolved-sideband limit correspond to a time-dependent carrier driving that interferes with the $\sigma^{\phi}$-force (see our discussion in Subsec.~\ref{sec:state-dep_forces} below), and thus compromises the geometric character of the gate and the achievable fidelities. 
 
 In this work, we show that  $\sigma^{\phi}_i\sigma^{\phi}_j$-gates with higher speeds and lower errors can be achieved by exploiting the {\it micromotion} of ion crystals, namely a periodic motion synchronous with the oscillations of the quadrupole potential that confines the ions in a Paul trap. We consider two different types of micromotion: {\it excess} and {\it intrinsic} micromotion. Excess micromotion can be described as a classical  driven motion  of the ions that lie off the r.f. null, either due to imperfections of the trap or to crystal configurations with equilibrium positions where the r.f. field does not vanish. The role of this excess micromotion on entangling-gate schemes has been considered previously, showing that {\it (i)} purposely-induced excess micromotion can be exploited to address different ions in a crystal  via  differential  Rabi frequencies of secular sidebands~\cite{mumotion_addressing}; {\it (ii)}  micromotion sidebands can be exploited to increase the gate speed with respect to schemes based on secular sidebands, in situations where the excess micromotion cannot be perfectly compensated~\cite{nist_MS}; and {\it (iii)} pulse sequences for  entangling gates based on either standard normal modes~\cite{mumotion_gates_2d, mumotion_gates_2d_bis} or solitonic vibrational excitations~\cite{mumotin_defects_gates}, can be designed even in the presence of the excess micromotion. With the exception of Ref.~\cite{mumotion_gates_2d}, the role of another type of micromotion in schemes of entangling gates, namely the intrinsic micromotion,  has remained largely unexplored.  Intrinsic micromotion corresponds to a quantum-mechanical driven motion synchronous with the r.f. frequency which cannot be compensated. Being quantum-mechanical, the intrinsic micromotion has a different impact on the gate schemes. In contrast to Ref.~\cite{mumotion_gates_2d}, where pulsed gate schemes are used to make the performance of the gate  equal to the ideal case where no micromotion is present,  we explore in this work the  possibility of  actively exploiting the intrinsic micromotion in order improve the gate performance, both in speed and  fidelity, beyond the values of the schemes where no micromotion is considered.

 This article is organised as follows. In Sec.~\ref{sec:mumotion}, we introduce the formalism that allows us to describe excess and intrinsic micromotion in generic ion crystals confined by Paul traps. This formalism is the starting point to develop  in Sec.~\ref{sec:entangling_gates} a general theory of laser-ion interactions in the regime of resolved sidebands in  presence of both excess and intrinsic micromotion. The expressions obtained are  then used to describe the main differences of the schemes that generate state-dependent dipole forces using bi-chromatic laser beams, either tuned to the secular or to the  micromotion sidebands. We also describe how these forces can be used to implement entangling gates, and   discuss the speed and fidelity limitations of various gate schemes, identifying a parameter regime where a gate improvement can be obtained by exploiting the intrinsic micromotion. In Sec.~\ref{sec:experimental_cons}, we discuss the possible experimental challenges in reaching such  parameter regime. Finally, we present our conclusions and outlook in Sec.~\ref{sec:conclusions}. 

\section{\bf Intrinsic and excess micromotion}

\label{sec:mumotion}

In this section, we start by reviewing the classical treatment of micromotion for a single trapped ion  in~\ref{sec:class_mumotion}. This will allow us to set the notation, and to explicitly define the notions of intrinsic and excess micromotion in ion traps. Additionally, it will provide some results that will be useful in the subsequent quantum-mechanical treatment in~\ref{sec:quant_mumotion}. The micromotion of a trapped-ion crystal is described in~\ref{sec:mumotion_crystals}, which shall be the starting point for the scheme of micromotion-enabled improvement of  quantum gates in the following section.  

\subsection{Classical treatment of  micromotion for a single trapped ion}
\label{sec:class_mumotion}

For the ease of exposition,  we focus in this section on the electric potential configuration and micromotion effects of an ion confined in a standard  linear Paul trap~\cite{rmp_leibfried}. We note that a similar analysis would apply to  segmented  linear traps~\cite{segmented_linear}, or to surface ion traps~\cite{surface}, which form a key central element in various scalable architectures  for QIP under development~\cite{ions_qc_scale}. At the end of this section, we will  comment on the analogies and possible differences for the micromotion in these other traps.  

We consider an ion of mass $M$ and charge $Q$, inside a standard linear Paul trap formed by {\it (i)} a pair of  end-caps separated by a distance $2z_0$ along the trap axis (i.e. $z$ axis), and connected to d.c. potentials $U_0$; {\it (ii)} four electrodes  separated from the axis by a distance $r_0$, and parallel to it, which are  connected in pairs to either a d.c. potential $V_0$, or an  a.c. potential $V_0\cos\Omega_{\rm rf} t$, where $\Omega_{\rm rf}$ is a fast r.f. frequency.  Accordingly, the ion is  subjected to an  oscillating quadrupole potential 
\beq
\label{eq:quad_potential}
V_{\rm q}=\frac{\kappa U_0}{2z_0^2}\bigg(\!2z^2-(x^2+y^2)\!\!\bigg)+\frac{ V_0\cos(\Omega_{\rm rf}t)}{2}\left(\!1+ \frac{1}{r_0^2}\left(x^2-y^2\right)\!\!\right),
\eeq
where $\kappa$ is a geometric factor that depends on the details of the electrodes. Here, we have assumed that the ions positions fulfill $|\boldsymbol{r}|\ll r_0,z_0$,  such that they lie  close to the trap axis and trap center. In this way, we are neglecting corrections to the quadrupole potential, such as as small component of the alternating r.f. field along the direction of the trap axis. 

In addition to the ideal quadrupole potential \eqref{eq:quad_potential},  there can be spurious potentials stemming from {\it (a)} potential variations due to patch effects, or to unevenly coated (charged) electrodes with elements (electrons)  coming from the oven (ionization process), and {\it (b)} asymmetries in the electrode impedances~\cite{excess_mumotion}. The former  leads to spurious d.c. fields ${\bf E}_{\rm dc}$ that displace the ions from the nodal line of the a.c. potential, whereas the latter induce small phase differences in the a.c. electrodes $\varphi_{\rm ac}$, which  give rise to an additional a.c. field. This field can be  approximated by that of a pair of parallel plates connected to potentials $\pm\half V_0\varphi_{\rm ac}\sin(\Omega_{\rm rf}t)$, and separated by $2r_0/\tilde{\alpha}$  with $\tilde{\alpha}$ being another geometric factor that depends on the trap configuration. These spurious effects thus lead to an additional potential
\beq
\label{eq:spur_potential}
V_{\rm s}=-\left({E}^x_{\rm dc}+\frac{V_0\varphi_{\rm ac}\tilde{\alpha}}{2r_0}\sin(\Omega_{\rm rf}t)\right)x-{E}^y_{\rm dc}y-{E}^z_{\rm dc}z.
\eeq

The classical equations of motion for the  ion correspond to a set of inhomogeneous Mathieu equations 
\beq
\label{eq:forced_mathieu}
\frac{{\rm d}^2r_\alpha}{{\rm d}\tau^2}+\big(a_\alpha-2q_\alpha\cos2\tau\big)r_\alpha=f_{\alpha}(\tau), \hspace{2ex}\alpha\in\{x,y,z\},
\eeq
where we have introduced the dimensionless time $\tau=\half\Omega_{\rm rf}t$, and the following dimensionless parameters
\beq
\label{eq:a,q}
a_x=a_y=-\frac{a_z}{2}=-\frac{4Q\kappa U_0}{Mz_0^2\Omega_{\rm rf}^2},\hspace{1ex} q_x=-q_y=-\frac{2QV_0}{Mr_0^2\Omega_{\rm rf}^2},\hspace{0.5ex}q_z=0.
\eeq
In addition, we get force terms in Eq.~\eqref{eq:forced_mathieu} due to the spurious potential~\eqref{eq:spur_potential},  namely
\beq
\label{eq:forces}
f_\alpha(\tau)=\frac{{4QE}^\alpha_{\rm dc}}{M\Omega_{\rm rf}^2}+\delta_{\alpha,x}\frac{2QV_0\varphi_{\rm ac}\tilde{\alpha}}{M r_0\Omega_{\rm rf}^2}\sin(2\tau),
\eeq
where we have used the Kronecker delta $\delta_{\alpha,\beta}$ in front of the contribution that stems from  the electron-impedance asymmetries, which leads to the small phase difference between the electrodes along the $x$-axis. The solution of these differential equations builds on the solution $r_\alpha^{\rm h}(\tau)$ to the homogeneous  Mathieu equation (i.e. $f_\alpha(\tau)=0$)~\cite{ab_stegun} by applying the method of variation of constants~\cite{boyce_di_prima}. Due to the periodicity of the equation, the solution can be expressed using the Floquet theorem as follows 
\beq
\label{eq:ind_solutions}
r^{\rm h}_\alpha(\tau)=\sum_{\frak{n}\in\mathbb{Z}}C_{2\frak{n}}^\alpha(A_\alpha\ee^{\ii(\beta_\alpha+2\frak{n})\tau}+B_\alpha\ee^{-\ii(\beta_\alpha+2\frak{n})\tau}),
\eeq
where $A_\alpha,B_\alpha$ are constants that depend on the initial conditions, $\beta_\alpha$ are the so-called characteristic exponents, and $C_{2\frak{n}}^\alpha$ are the Floquet coefficients. By substitution, one finds that these coefficients fulfill a recursion relation
\beq
\label{eq:single_ion_recursion}
C^\alpha_{2\frak{n}+2}-D^\alpha_{2\frak{n}}C^\alpha_{2\frak{n}}+C^\alpha_{2\frak{n}-2}=0,\hspace{2ex} D^\alpha_{2\frak{n}}=\frac{a_\alpha-(\beta_\alpha+2\frak{n})^2}{q_\alpha}.
\eeq
In typical experimental realizations, the parameters~\eqref{eq:a,q} fulfill 
\beq 
\label{eq:constraint}
a_\alpha,q_\alpha^2\ll1,
\eeq
 and this allows one to solve the above recursion to the desired order of accuracy. To the lowest-possible order, one finds 
\beq
\beta_\alpha=\sqrt{a_\alpha+\half q_\alpha^2},\hspace{2ex} C_{\pm 2\ell}^\alpha=\frac{(-1)^\ell q_{\alpha}^\ell C_{0}^\alpha}{4^\ell((\ell-1)!)^2},
\eeq
where we have introduced a positive integer $\ell$ to label the different harmonics. Note that  $C_{- 2\ell}^\alpha\neq C_{2\ell}^\alpha$ for general parameters. However, this difference can be neglected 
to leading order in the small parameters~\eqref{eq:constraint}. Imposing that $r_\alpha(0)=r_\alpha^0,{\rm d}r_\alpha(\tau)/{\rm d}\tau|_{\tau=0}=0$, and $C_0^\alpha=1$, we find that the homogeneous solution describing the motion of an ion inside an ideal Paul trap is 
\beq
\label{eq:hom_solution}
r^{\rm h}_\alpha(t)=\frac{r_\alpha^0}{\xi_\alpha} \cos(\omega_\alpha t)\left(1+\sum_{\ell\geq 1}\frac{(-1)^\ell 2q_{\alpha}^\ell}{4^\ell((\ell-1)!)^2}\cos(\ell\Omega_{\rm rf}t)\right),
\eeq
where we have introduced the so-called secular frequencies
\beq
\label{eq:sec_freq}
\omega_\alpha =\frac{\Omega_{\rm rf}}{2}\beta_\alpha,
\eeq
which are much smaller than the trap r.f. frequency $\omega_\alpha\ll \Omega_{\rm rf}$. We have also introduced the parameter 
\beq
\label{eq:norm_xi}
\xi_\alpha=1+\sum_{\ell\geq 1}(-1)^\ell\frac{ 2q_{\alpha}^\ell}{4^\ell((\ell-1)!)^2}.
\eeq

We can rewrite Eq.~\eqref{eq:hom_solution} as $r^{\rm h}_\alpha(t)=r^{\rm sec}_\alpha(t)+r^{\rm in}_\alpha(t)$, such that the ion in an ideal Paul trap  displays slow oscillations at the secular frequency  described by
\beq
\label{eq:sec_solution}
r^{\rm sec}_\alpha(t)=\frac{r_\alpha^0}{\xi_\alpha} \cos(\omega_\alpha t),
\eeq
accompanied by smaller and faster oscillations synchronous with the a.c. potential 
\beq
\label{eq:in_solution}
r^{\rm in}_\alpha(t)=r^{\rm sec}_\alpha(t)\left(\sum_{\ell\geq 1}\frac{(-1)^\ell 2q_{\alpha}^\ell}{4^\ell((\ell-1)!)^2}\cos(\ell\Omega_{\rm rf}t)\right).
\eeq
 These smaller oscillations are referred to as {\it micromotion}, and occur roughly at multiples of the r.f. frequency $\ell \Omega_{\rm rf}$ (i.e. micromotion sidebands). To distinguish this behavior from the one stemming from the spurious potential~\eqref{eq:spur_potential}, these fast oscillations $r^{\rm in}_\alpha(t)$ are sometimes referred to as {\it intrinsic micromotion}~\cite{mu_motion_compensation}, to highlight the fact  that such a motion is intrinsic to the  oscillating quadrupole of an  Paul trap, even for an ideal trap design without any imperfection~\eqref{eq:spur_potential}.

The solution to the forced Mathieu equation~\eqref{eq:forced_mathieu}  can be found using the method of variation of constants with the two independent solutions associated to Eq~\eqref{eq:ind_solutions}, namely $r_\alpha^{\rm h}(t)=A_\alpha r_{\alpha,1}(t)+B_\alpha r_{\alpha,2}(t)$. The complete solution is  
\beq
\label{eq:full_solution}
r_\alpha(t)=r^{\rm sec}_\alpha(t)+r^{\rm in}_\alpha(t)+r^{\rm ex}_\alpha(t),
\eeq
 where the additional part due to the spurious potential is 
\beq
r^{\rm ex}_\alpha(t)=\!\!\int\!{\rm d}\tau'\frac{\left(r_{\alpha,2}(\tau)r_{\alpha,1}(\tau')-r_{\alpha,1}(\tau)r_{\alpha,2}(\tau')\right)f_\alpha(\tau')}{\mathbb{W}_\alpha(\tau')},
\eeq
and $\mathbb{W}_\alpha(\tau')=r_{\alpha,1}(\tau'){\rm d} r_{\alpha,2}(\tau')/{\rm d}\tau'-r_{\alpha,2}(\tau'){\rm d} r_{\alpha,1}(\tau')/{\rm d}\tau'$ is the Wronskian of the two solutions, which can be shown to be constant in this case $\mathbb{W}(\tau')=-2\ii\beta_\alpha$. When performing the integrals, we keep only the slowly-varying terms, which give rise to the leading-order solution
\beq
\label{eq:part_solution}
r^{\rm ex}_\alpha(t)= r_\alpha^{\rm driv}(t)\left(1+\sum_{\ell\geq 1}\frac{(-1)^\ell 2q_{\alpha}^\ell}{4^\ell((\ell-1)!)^2}\cos(\ell\Omega_{\rm rf}t)\right),
\eeq
where we have introduced the following driven amplitude
\beq
\label{eq:driven_motion}
 r_\alpha^{\rm driv}(t)=\frac{QE_{\rm dc}^\alpha}{M\omega_\alpha^2}+\delta_{\alpha,x}\frac{q_xr_0\varphi_{\rm ac}\tilde{\alpha}}{4}\sin(\Omega_{\rm rf}t).
\eeq
We thus observe that the spurious potential~\eqref{eq:spur_potential} induces  a driven motion~\eqref{eq:part_solution} that is also synchronous with the r.f. frequency, and is thus another type of micromotion. Since it is not linked to the secular motion, and can only be reduced by compensating the spurious potential terms~\eqref{eq:spur_potential}, this motion is usually referred to as {\it excess micromotion}~\cite{excess_mumotion}. Along this text, we will use the wording micromotion compensation to refer to the compensation of the stray fields that produce excess micromotion.

As a consistency check, we note that to  linear order in $q_\alpha$, the  complete solution~\eqref{eq:full_solution}  built from Eqs.~\eqref{eq:sec_solution},~\eqref{eq:in_solution} and~\eqref{eq:part_solution} coincides with the solution presented in~\cite{excess_mumotion}, which includes the secular motion and the first micromotion sideband. The higher-order powers of $q_\alpha$ allow us to account for all higher micromotion sidebands. Note also that the intrinsic~\eqref{eq:in_solution} and excess~\eqref{eq:part_solution} micromotion only occur in those trap axes where $q_\alpha\neq 0$. According to Eq.~\eqref{eq:a,q}, micromotion in an ideal  linear Paul trap only occurs in the transverse directions, as there is no r.f. field along the axial direction, such that $q_z=0$. However, for realistic experimental conditions that depart from this ideal case, there might also be axial micromotion $q_z\neq0$, as occurs for instance in segmented linear traps~\cite{mu_motion_compensation}. Accordingly, we will consider the most general case, and allow for micromotion in all possible directions $q_\alpha\neq 0,\forall\alpha\in\{x,y,z\}$. The particular microscopic expression of these parameters will generally differ from Eq.~\eqref{eq:a,q}, and depend on specific details of the trap. For the excess micromotion~\eqref{eq:part_solution}, the driven amplitude $r_\alpha^{\rm driv}(t)$ will differ from the ideal case~\eqref{eq:driven_motion}, and  also depend on specific details of the trap. However, one can  treat the micromotion  generically using  Eqs.~\eqref{eq:sec_solution},~\eqref{eq:in_solution} and~\eqref{eq:part_solution}, with generic  parameters $ r_\alpha^{\rm driv}(t), a_\alpha,q_\alpha$ only restricted to fulfill Eq.~\eqref{eq:constraint}.   

\subsection{Quantum-mechanical treatment of micromotion for a single trapped ion}
\label{sec:quant_mumotion}

Since the ultimate goal of this work is to exploit the micromotion to improve   phonon-mediated quantum logic gates between distant trapped-ion qubits, a full quantum-mechanical treatment of the secular vibrations and the intrinsic/excess micromotion in a trapped-ion crystal will be required. A detailed quantum-mechanical treatment of the secular vibrations and intrinsic micromotion for a single trapped ion has been described in~\cite{rmp_leibfried} using a formalism  based on  quantum-mechanical constants of motion~\cite{mu_motion_glauber}. We now use  this formalism to generalize the description  to  situations where  excess micromotion of a single trapped ion is also present. 

The quantum-mechanical Hamiltonian of the ion inside the Paul trap can be described as
\beq
\label{eq:H_mumotion}
H=\sum_{\alpha}\left(\frac{1}{2M}\hat{p}_{\alpha}^2+\frac{1}{2}K_\alpha(t) \hat{r}_{\alpha}^2-MF_{\alpha}(t)\hat{r}_\alpha\right),
\eeq
where we have promoted the position and momentum to quantum-mechanical operators fulfilling $[\hat{r}_\alpha,\hat{p}_{\beta}]=\ii\delta_{\alpha,\beta}$. Here, we have introduced a time-dependent spring constant 
\beq 
\label{eq:spring_constants}
K_\alpha(t)=\frac{M}{4} \Omega_{\rm rf}^2\big(a_\alpha-2q_\alpha\cos\Omega_{\rm rf}t\big),
\eeq
 and used the time-dependent forces~\eqref{eq:forces} caused by the deviations~\eqref{eq:spur_potential} from an ideal Paul trap, transformed back into real time 
 \beq
 \label{eq:spurious_forces}
 F_{\alpha}(t)=\fourth\Omega^2_{\rm rf}f_\alpha\left(\half\Omega_{\rm rf}t\right).
 \eeq
  The Heisenberg equations of motion for this Hamiltonian lead to a quantum-mechanical version of the classical forced Mathieu equations~\eqref{eq:forced_mathieu} for the position operator, namely 
\beq
\label{eq:forced_mathieu_quantum}
\frac{{\rm d}^2\hat{r}_\alpha(t)}{{\rm d}t^2}+\frac{K_\alpha(t)}{M}\hat{r}_\alpha(t)=F_{\alpha}(t), \hspace{2ex}\alpha\in\{x,y,z\}.
\eeq
We now construct an operator constant of motion by combining the position operator $\hat{r}_{\alpha} (t)$ fulfilling Eq.~\eqref{eq:forced_mathieu_quantum}, with a mode function $u_{\alpha}(t)$ that
evolves according to the solution of the homogeneous classical Mathieu equation~\eqref{eq:ind_solutions}, but with initial conditions $u_\alpha(0)=1,{\rm d}u_\alpha(t)/{\rm d}t|_{t=0}=\ii\omega_\alpha$, and $C_0^\alpha=1$. This  generalises the standard mode function $u_{\alpha}^{\rm st}(t)=\ee^{\ii\omega_\alpha t}$ that appears in the Heisenberg picture of a time-independent harmonic oscillator of frequency $\omega_\alpha$, and can be expressed as follows
\beq
\label{eq:mod_solutions}
u_\alpha(t)=\frac{\ee^{\ii\omega_\alpha t}}{\xi_\alpha}\left(1+\sum_{\ell\geq 1}\frac{(-1)^\ell 2q_{\alpha}^\ell}{4^\ell((\ell-1)!)^2}\cos(\ell\Omega_{\rm rf}t)\right),
\eeq
where we   used  the secular frequencies~\eqref{eq:sec_freq} and the normalization~\eqref{eq:norm_xi}. The operator constant of motion is built from the Wronskian of the position operator and the mode function $\widehat{\mathbb{W}}_\alpha(t)=u_\alpha(t){\rm d}\hat{r}_\alpha(t)/{\rm d}t-\hat{r}_\alpha(t){\rm d}u_\alpha(t)/{\rm d}t$, namely
\beq
\label{eq:cont_operator}
a_\alpha(t)=\ii\sqrt{\frac{M}{2\omega_\alpha}}\left(\widehat{\mathbb{W}}_\alpha(t)-\int_0^t{\rm d}t'u_\alpha(t')F_\alpha(t')\right),
\eeq
which fulfills $a_\alpha(t)=a_\alpha$, where 
\beq
a_\alpha=\sqrt{\frac{M\omega_\alpha}{2}}\left(\hat{r}_\alpha+\frac{\ii}{M\omega_\alpha}\hat{p}_\alpha\right)
\eeq
is the standard annihilation operator of a harmonic oscillator vibrating at the secular frequency. Using these expressions, and keeping once more the slowly-varying terms under the integral of Eq.~\eqref{eq:mod_solutions}, we find that the quantum-mechanical position operator can be expressed as follows
\beq
\label{eq:full_solution_quantum}
\hat{r}_\alpha(t)=\hat{r}^{\rm sec}_\alpha(t)+\hat{r}^{\rm in}_\alpha(t)+r^{\rm ex}_\alpha(t)\widehat{\mathbb{I}}.
\eeq Here, the secular-motion position operator is given by
\beq
\label{eq:sec_motion_operator}
\hat{r}^{\rm sec}_\alpha(t)=\frac{1}{\sqrt{2M\omega_\alpha}}\frac{1}{\xi_\alpha}\left(a_{\alpha}^\dagger\ee^{\ii\omega_\alpha t}+a_{\alpha}^{\phantom{\dagger}}\ee^{-\ii\omega_\alpha t}\right).
\eeq 
The intrinsic micromotion operator can be expressed as 
\beq
\label{eq:in_mumotion_operator}
\hat{r}^{\rm in}_\alpha(t)=\hat{r}^{\rm sec}_\alpha(t)\left(\sum_{\ell\geq 1}\frac{(-1)^\ell 2q_{\alpha}^\ell}{4^\ell((\ell-1)!)^2}\cos(\ell\Omega_{\rm rf}t)\right),
\eeq 
and is thus again proportional to the secular motion, as occurred in the classical case~\eqref{eq:in_solution}. Since the secular motion is expressed in terms of quantum-mechanical creation-annihilation operators, the intrinsic micromotion can be argued to be a quantum-mechanical motion synchronous with the r.f. oscillating field, as we advanced in the introduction of this work. As a consistency check, we note that to linear order in $q_\alpha$, we recover the expressions  described in~\cite{rmp_leibfried}. 

Finally, Eq.~\eqref{eq:full_solution_quantum} also includes the effects of  excess micromotion in the position operator, which are proportional to the identity operator in the vibrational Hilbert space $\widehat{\mathbb{I}}$. As expected from the forces in Eq.~\eqref{eq:H_mumotion},  the excess micromotion corresponds to a simple displacement over the position operator $\hat{r}^{\rm sec}_\alpha(t)\to \hat{r}^{\rm sec}_\alpha(t)+ r_\alpha^{\rm driv}(t)\widehat{\mathbb{I}}$, the magnitude of   which coincides exactly with the classical driven amplitude~\eqref{eq:driven_motion}. One thus finds that the deviations from an ideal Paul trap affect the quantum-mechanical position operator~\eqref{eq:full_solution_quantum} through the classical expression $r^{\rm ex}_\alpha(t)$ of the excess micromotion~\eqref{eq:part_solution}. Accordingly, the excess micromotion can be considered as a classical driven motion that can indeed be compensated by minimizing the spurious terms~\eqref{eq:spur_potential}, in contrast to the intrinsic micromotion.

\subsection{ Quantum-mechanical treatment of micromotion in a trapped-ion crystal}
\label{sec:mumotion_crystals}

The standard treatment of phonons in solids  considers small quantized displacements of the ions  $\delta \hat{r}_{i,\alpha}$ with respect to an underlying  Bravais lattice $r_{i,\alpha}^0$, namely 
$\hat{r}_{i,\alpha}(t)=r_{i,\alpha}^0+\delta \hat{r}_{i,\alpha}$, where $i\in\{1,\dots,N\}$ labels a particular ion. The collective modes of vibration, whose quantum-mechanical excitations lead to the aforementioned phonons, are usually obtained in the harmonic approximation by expanding the  inter-ionic potential to second order in the  displacements, and  diagonalising the resulting quadratic Hamiltonian~\cite{feynman_stat}. For a collection of $N$ ions inside a Paul trap,  an analogous treatment exists in the
so-called pseudo-potential approximation, which assumes that the ions are effectively trapped by a time-independent quadratic potential with  secular trap frequencies~\eqref{eq:sec_freq}. The main difference with respect to a solid is that the equilibrium positions  $r_{i\alpha}^0$ do not correspond to a Bravais lattice, but instead form an inhomogeneous array known as a Coulomb crystal~\cite{normal_modes}. This approximation, however, does not include possible effects of micromotion in the ion crystal.  

A careful classical treatment of the  crystal micromotion~\cite{mumotion_crystals} has recently shown that it can have non-trivial effects, such as a renormalisation of the normal-mode  frequencies in planar crystals~\cite{mu_motion_modes_exp}. We now present a detailed quantum-mechanical treatment of both the intrinsic and excess micromotion in ion crystals, which combines the techniques presented in Sec.~\ref{sec:class_mumotion} with the formalism in~\cite{mumotion_crystals} to describe the effect of micromotion on the classical crystal, and then generalises Sec.~\ref{sec:quant_mumotion}  to describe quantum-mechanically the effect of micromotion on the phonons of the ion crystal.

To incorporate the different types of micromotion introduced above, we generalise the quantum-mechanical Hamiltonian~\eqref{eq:H_mumotion} to a system of $N$ ions confined by an oscillating quadrupole
\beq
\label{eq:H_mumotion_crystal}
H=\!\sum_{i,\alpha}\left(\!\frac{\hat{p}_{i,\alpha}^2}{2M}+\frac{1}{2}K_\alpha(t) \hat{r}_{i,\alpha}^2-MF_{i,\alpha}(t)\hat{r}_{i,\alpha}\right)+\sum_{ i,j}\frac{\tilde{Q}^2}{2|\hat{\boldsymbol{ r}}_{i}-\hat{\boldsymbol{ r}}_{j}|},
\eeq
where we have introduced  vectorial operators defined in terms of the unit cartesian vectors  $\hat{\boldsymbol{ r}}_{i}=\sum_\alpha\hat{{ r}}_{i,\alpha}{\bf e}_\alpha$,   and   the position-momentum operators now fulfill  $[\hat{r}_{i,\alpha},\hat{p}_{j,\beta}]=\ii\delta_{\alpha,\beta}\delta_{i,j}$. Here,   $\tilde{Q}^2={Q}^2/4\pi\epsilon_0$ eases notation,  and  we have used the  spring constants~\eqref{eq:spring_constants} and time-dependent forces~\eqref{eq:spurious_forces} introduced above, allowing the spurious d.c. fields in Eq.~\eqref{eq:forces} to be inhomogeneous along the crystal. The Heisenberg equations of motion lead to a system of equations
\beq
\label{eq:forced_mathieu_quantum_crystal}
\frac{{\rm d}^2\hat{r}_{i,\alpha}}{{\rm d}t^2}+\frac{K_\alpha(t)}{M}\hat{r}_{i,\alpha}-\tilde{Q}^2\sum_{j\neq i}\frac{\hat{r}_{i,\alpha}-\hat{r}_{j,\alpha}}{|\hat{\boldsymbol{r}}_{i}-\hat{\boldsymbol{r}}_{j}|^3}=F_{i,\alpha}(t).
\eeq

Paralleling the standard treatment of phonons in solids, we substitute 
\beq
\label{eq:harmonic_approx}
\hat{r}_{i,\alpha}\to {r}^{0}_{i,\alpha}(t)\widehat{\mathbb{I}}+ \delta \hat{r}_{i,\alpha},
\eeq
 where $ {r}^{0}_{i,\alpha}(t)$
 are the equivalent of the equilibrium positions in solids, which become time-dependent quantities in the presence of micromotion (i.e. breathing crystal),  and $\delta \hat{r}_{i,\alpha}$ are the small quantized vibrations around such a  breathing crystal. When these vibrations are sufficiently small, the equations~\eqref{eq:forced_mathieu_quantum_crystal} decouple into {\it (i)} a classical system of differential equations for the coordinates of the breathing crystal, and {\it (ii)} a linear system of equations for the quantum-mechanical displacements. 

Let us focus on {\it (i)}, and  rescale the time $\tau=\half\Omega_{\rm rf}t$, such that the time-periodic breathing crystal fulfils
\beq
\label{eq:mathieu_time_crystal}
\frac{{\rm d}^2r^0_{i,\alpha}}{{\rm d}\tau^2}+\big(a_\alpha-2q_\alpha\cos2\tau\big)r^0_{i,\alpha}-\frac{4\tilde{Q}^2}{M\Omega_{\rm rf}^2}\sum_{j\neq i}\frac{{r}^0_{i,\alpha}-{r}^0_{j,\alpha}}{|{\boldsymbol{r}}^0_{i}-\boldsymbol{r}^0_{j}|^3}=0.
\eeq
 These differential equations correspond to a system of coupled Mathieu equations~\eqref{eq:forced_mathieu} and, inspired by the previous section,  we thus propose a Floquet-type ansatz the form of Eq.~\eqref{eq:ind_solutions}, namely 
\beq
\label{eq:ind_solutions_crystal}
r^{\rm 0}_{i,\alpha}(\tau)=\sum_{\frak{n}\in\mathbb{Z}}C_{2\frak{n},i}^\alpha(A_\alpha\ee^{\ii(\beta_\alpha+2\frak{n})\tau}+B_\alpha\ee^{-\ii(\beta_\alpha+2\frak{n})\tau}),
\eeq
where  $A_\alpha,B_\alpha$ are constants that depend on the initial conditions, $\beta_\alpha$ is the so-called characteristic exponent, and $C_{2\frak{n},i}^\alpha$ are the Floquet coefficients. The breathing crystal corresponds to a classical solution of the type~\eqref{eq:ind_solutions_crystal} synchronous with the r.f. potential, i.e. $\beta_\alpha=0$, and can also be considered as part of the excess micromotion due to ion positions lying off the r.f. null. By substituting this expression~\eqref{eq:ind_solutions_crystal} in Eq.~\eqref{eq:mathieu_time_crystal}, one observes that the Coulomb repulsion can introduce higher harmonics of the r.f. frequency. For the linear ion crystals of interest to our purposes,   these effects are absent in the relevant parameter regime~\eqref{eq:constraint}, where the Floquet coefficients  fulfill a system of coupled recursion relations
\beq
\label{eq:recursion_eqs_crystal}
C^\alpha_{2\frak{n}+2,i}-D_{2\frak{n}}^\alpha C^\alpha_{2\frak{n},i}-\sum_{j\neq i}\frac{\zeta(C^\alpha_{2\frak{n},i}-C^\alpha_{2\frak{n},j})}{\big[\sum_\alpha(C^\alpha_{0,i}-C^\alpha_{0,j})^2\big]^{\frac{3}{2}}}+C^\alpha_{2\frak{n}-2,i}=0.
\eeq
Here, we have used the same notation as in the recursion relation for a single ion~\eqref{eq:single_ion_recursion}, $D^\alpha_{2\frak{n}}=(a_\alpha-4\frak{n}^2)/q_\alpha$, and introduced the parameter $\zeta=4\tilde{Q}^2/M\Omega_{\rm rf}^2q_\alpha$. To the lowest possible order in~\eqref{eq:constraint}, one finds that all the Floquet coefficients 
\beq
 C_{\pm 2\ell,i}^\alpha=\frac{(-1)^\ell q_{\alpha}^\ell C_{0,i}^\alpha}{4^\ell((\ell-1)!)^2}
\eeq
are   expressed in terms of the time-independent one $C_{0,i}^\alpha$. This coefficient is  in turn  determined by   the solutions of the following system of algebraic equations
\beq
\label{eq:crystal_eq_positions}
M\omega_\alpha^2C^\alpha_{0,i}-\sum_{j\neq i}\frac{\tilde{Q}^2(C^\alpha_{0,i}-C^\alpha_{0,j})}{\big[\sum_\alpha(C^\alpha_{0,i}-C^\alpha_{0,j})^2\big]^{\frac{3}{2}}}=0,
\eeq
where we have made use of the secular trapping frequencies introduced in Eq.~\eqref{eq:sec_freq}. Let us note that these equations display a clear  competition between the harmonic trapping and the Coulomb repulsion, and coincide with those that determine the equilibrium positions of the ion crystal in  the pseudo-potential approximation~\cite{normal_modes}. Therefore, we 
 shall  denote the solutions as $r_{i,\alpha}^{\rm eq}=C^\alpha_{0,i}$, which can be found numerically.

Since we are interested in the linear-trap configuration $\omega_{z}\ll\omega_x,\omega_y$, where $ r_{i,\alpha}^{\rm eq}=z_i^0\delta_{\alpha,z}$, and the oscillating quadrupole has no effect along the trap axis  of an ideal Paul trap $q_z=0$; we find that $ C_{\pm 2\ell,i}^\alpha=0$, $\forall \ell\geq 1$ at this leading order. In fact, only terms at a higher-power of the non-vanishing parameters $q_x,q_y$ can lead to corrections~\cite{mumotion_crystals}, but these are negligible in the regime of Eq.~\eqref{eq:constraint}. 
 Accordingly, 
the time-periodic breathing  crystal~\eqref{eq:ind_solutions_crystal} in an ideal Paul trap corresponds to   a static  Coulomb crystal 
\beq
\label{eq:linear_periodic_crystal}
\boldsymbol{r}_{i}^{0}(t)=z_i^0{\bf e}_{z}.
\eeq
In a segmented linear trap, where residual axial micromotion may exist $0<q_z\ll q_x,q_y$, one would still obtain a static crystal to leading order. Conversely, for crystalline solutions where ions lie off the trap axis,  the higher-order harmonics  introduced by the Coulomb interaction in Eq.~\eqref{eq:mathieu_time_crystal} for a breathing crystal with $ q_x,q_y>0$ must be considered in detail for an accurate description~\cite{mumotion_crystals,mu_motion_modes_exp}.

Given this solution~\eqref{eq:linear_periodic_crystal}, we can now turn our attention onto {\it (ii)}, namely the quantum-mechanical displacements about the crystal. After linearization, one can show that the corresponding operators evolve according to a system of forced Mathieu equations, similar to the single-ion case~\eqref{eq:forced_mathieu_quantum}, but now coupled via the linearised Coulomb interaction
\beq
\frac{{\rm d}^2}{{\rm d}t^2}\delta\hat{r}_{i,\alpha}(t)+\sum_{j}\frac{\mathbb{K}^\alpha_{ij}(t)}{M}\delta\hat{r}_{j,\alpha}(t)=F_{i,\alpha}(t), 
\eeq
where we have now time-dependent spring constants that couple distant ions
\beq
\label{eq:coll_spring_cte}
\mathbb{K}^\alpha_{ij}(t)=\delta_{i,j}K_\alpha(t)+ \mathbb{V}^\alpha_{ij}(t).
\eeq
Here, we have used the single-ion spring constants~\eqref{eq:spring_constants}, and introduced the matrix of Coulomb-mediated couplings
\beq
\begin{split}
 \mathbb{V}^\alpha_{ij}(t)&=\frac{ (1-\delta_{i,j})\tilde{Q}^2}{|{\boldsymbol{r}}^0_{i}(t)-\boldsymbol{r}^0_{j}(t)|^3}(\delta_{\alpha,x}+\delta_{\alpha,y}-2\delta_{\alpha,z})\\
 &-\sum_{\ell\neq i}\frac{\delta_{i,j} \tilde{Q}^2}{|{\boldsymbol{r}}^0_{i}(t)-\boldsymbol{r}^0_{\ell}(t)|^3}(\delta_{\alpha,x}+\delta_{\alpha,y}-2\delta_{\alpha,z}).
 \end{split}
\eeq

For the linear crystals~\eqref{eq:linear_periodic_crystal} that concern us in this work,  this coupling matrix becomes time-independent  $\mathbb{V}^\alpha_{ij}(t)= \mathbb{V}^\alpha_{ij}$, and the system of differential equations can be decoupled by a single orthogonal transformation, in analogy to the standard theory of phonons in solids~\cite{feynman_stat}. We thus introduce the following normal-mode operators
\beq
 \hat{R}_{m,\alpha}(t)=\sum_i\mathcal{M}_{i,m}^\alpha\delta\hat{r}_{i,\alpha}(t), \hspace{2ex} \hat{P}_{m,\alpha}(t)=\sum_i\mathcal{M}_{i,m}^\alpha\hat{p}_{i,\alpha}(t),
\eeq
where the orthogonal matrix is determined by diagonalizing the matrix of Coulomb-mediated couplings
\beq
\sum_{i,j}\mathcal{M}_{i,n}^\alpha\mathbb{V}^\alpha_{ij}\mathcal{M}_{j,m}^\alpha=\delta_{n,m}\mathcal{V}^\alpha_{m},
\eeq
where we have introduced the eigenvalues $\mathcal{V}^\alpha_{m}$.
Using the orthogonality of the transformation, we find a set of decoupled forced Mathieu equations for the normal-mode  operators
\beq
\label{eq:normal_mode_Mathieu}
\frac{{\rm d}^2}{{\rm d}t^2}\hat{R}_{m,\alpha}(t)+\frac{\kappa^\alpha_{m}(t)}{M}\hat{R}_{m,\alpha}(t)=\frak{F}_{m,\alpha}(t).
\eeq
Here,  the eigenvalues of the spring-coupling matrix 
\beq
\kappa^\alpha_{m}(t)=K_{\alpha}(t)+\mathcal{V}_m^\alpha
\eeq
 inherit the time-dependence via the  single-ion spring constants~\eqref{eq:spring_constants}, and we have introduced   forces that tend to displace the ions along the normal-mode directions 
\beq
\frak{F}_{m,\alpha}(t)=\sum_i\mathcal{M}^\alpha_{i,m}\tilde{F}_{i,\alpha}(t).
\eeq
Hence, we have reduced the dynamics of the small quantum displacements about the  crystalline solution into $3N$ instances of the single-ion problem~\eqref{eq:forced_mathieu_quantum}. We must thus find 3$N$ operators that are constants of motion, which requires finding a set of normal mode functions $u_{m,\alpha}(t)$ that are solutions of the homogeneous Mathieu equations~\eqref{eq:normal_mode_Mathieu}, namely
\beq
u_{m,\alpha}(t)=\!\!\sum_{\frak{n}\in\mathbb{Z}}\!C_{2\frak{n},m}^\alpha(A_{m,\alpha}\ee^{\ii(\beta_{m,\alpha}+2\frak{n})\frac{\Omega_{\rm rf}t}{2}}+B_{m,\alpha}\ee^{-\ii(\beta_{m,\alpha}+2\frak{n})\frac{\Omega_{\rm rf}t}{2}}).
\eeq
This is the generalization of Eq.~\eqref{eq:ind_solutions} with constants $A_{m,\alpha},B_{m,\alpha}$  that depend on the initial conditions,  characteristic exponents for each normal mode $\beta_{m,\alpha}$, and  Floquet coefficients $C_{2\frak{n},m}^\alpha$. We impose the  initial conditions $u_{m,\alpha}(0)=1, {\rm d}u_{m,\alpha}(t)/{\rm d}t|_{t=0}=\ii\omega_{m,\alpha}$, and $C_{0,m}^\alpha=1$, such that the mode functions   to leading order in~\eqref{eq:constraint} can be expressed as
\beq
\label{eq:normal_mod_solutions}
u_{m,\alpha}(t)=\frac{\ee^{\ii\omega_{m,\alpha} t}}{\xi_\alpha}\left(1+\sum_{\ell\geq 1}\frac{(-1)^\ell 2q_{\alpha}^\ell}{4^\ell((\ell-1)!)^2}\cos(\ell\Omega_{\rm rf}t)\right),
\eeq
with the normalization constant defined in Eq.~\eqref{eq:norm_xi}, and the normal-mode secular frequencies
\beq
\label{eq:sec_freq_crystal}
\omega_{m,\alpha}=\sqrt{\omega_\alpha^2+\mathcal{V}_m^\alpha},
\eeq
where we have used  the secular frequency in Eq.~\eqref{eq:sec_freq}. 

Given these normal-mode functions, one can obtain the Wronskian and the  constants of motion through a straightforward generalisation of Eq.~\eqref{eq:cont_operator} by introducing the annihilation operators for each collective vibrational mode
\beq
a_{m,\alpha}=\sqrt{\frac{M\omega_{m,\alpha}}{2}}\left( \hat{R}_{m,\alpha}+\frac{\ii}{M\omega_{m,\alpha}}\hat{P}_{m,\alpha}\right).
\eeq
Therefore, the analogue of Eq.~\eqref{eq:full_solution_quantum} for the quantum-mechanical treatment of  micromotion in a trapped-ion crystal can be expressed as
\beq
\label{eq:full_solution_quantum_crystal}
\hat{r}_{i,\alpha}(t)=r_{i,\alpha}^0(t)\widehat{\mathbb{I}}+\delta\hat{r}^{\rm sec}_{i,\alpha}(t)+\delta\hat{r}^{\rm in}_{i,\alpha}(t)+ r^{\rm ex}_{i,\alpha}(t)\widehat{\mathbb{I}}.
\eeq 
Here, the secular-motion position operator is given by
\beq
\label{eq:sec_motion_operator_crystal}
\delta\hat{r}^{\rm sec}_{i,\alpha}(t)=\sum_{m}\frac{\mathcal{M}_{i,m}^\alpha}{\sqrt{2M\omega_{m,\alpha}}}\frac{1}{\xi_\alpha}\left(a_{m,\alpha}^\dagger\ee^{\ii\omega_{m,\alpha} t}+a_{m,\alpha}^{\phantom{\dagger}}\ee^{\ii\omega_{m,\alpha} t}\right),
\eeq 
and the intrinsic micromotion operator can be expressed as 
\beq
\label{eq:in_mumotion_operator_crystal}
\delta\hat{r}^{\rm in}_{i,\alpha}(t)=\delta \hat{r}^{\rm sec}_{i,\alpha}(t)\left(\sum_{\ell\geq 1}\frac{(-1)^\ell 2q_{\alpha}^\ell}{4^\ell((\ell-1)!)^2}\cos(\ell\Omega_{\rm rf}t)\right).
\eeq 
The excess micromotion in Eq.~\eqref{eq:full_solution_quantum_crystal} is expressed in terms of  the identity operator in the vibrational Hilbert space $\widehat{\mathbb{I}}$, and  
\beq
\label{eq:part_solution_crystal}
 r^{\rm ex}_{i,\alpha}(t)= r_{i,\alpha}^{\rm driv}(t)\left(1+\sum_{\ell\geq 1}\frac{(-1)^\ell 2q_{\alpha}^\ell}{4^\ell((\ell-1)!)^2}\cos(\ell\Omega_{\rm rf}t)\right),
\eeq
where we have introduced the generic site-dependent  amplitude $ r_{i,\alpha}^{\rm driv}(t)$. For the standard Paul trap, this can be obtained from Eq.~\eqref{eq:driven_motion} by considering inhomogeneous spurious fields
\beq
\label{eq:driven_motion_crystal}
 r_{i,\alpha}^{\rm driv}(t)=\frac{QE_{{\rm dc},i}^\alpha}{M\omega_\alpha^2}+\delta_{\alpha,x}\frac{q_xr_0\varphi_{\rm ac}\tilde{\alpha}}{4}\sin(\Omega_{\rm rf}t).
\eeq
For other situations, such as those arising for segmented traps, this  amplitude will depend on the specific trap details. 

In this way, we have presented   a detailed quantum-mechanical description of the effects of intrinsic and excess micromotion in a linear crystal of trapped ions. The results in Eqs.~\eqref{eq:full_solution_quantum_crystal}-\eqref{eq:driven_motion_crystal}  will be the starting point for the scheme of micromotion-enabled entangling gates in the following section.  Our formalism can be extended to planar crystals, although one has to consider the breathing of the crystal instead of Eq.~\eqref{eq:linear_periodic_crystal}, and how this can lead to micromotion-induced corrections of the secular vibrations of the planar crystal.

\section{\bf Entangling gates based on micromotion sidebands}
\label{sec:entangling_gates}

In this section, we start in~\ref{sec:mumotion_sidebands} by discussing the effects of micromotion in the theory of light-matter interactions for a set of laser beams addressed to a particular electronic transition of the ions. We then describe how to implement state-dependent forces by combining pairs of laser beams in~\ref{sec:state-dep_forces}, and discuss the role of intrinsic/excess micromotion, paying special attention to the contribution of the often-neglected carrier excitations.  In~\ref{sec:gates}, we start by reviewing the schemes for entangling gates based on secular state-dependent $\sigma^\phi$-forces~\cite{ms_gates,ms_gates_nist,innsbruck_MS,oxford_MS,nist_MS}, and discussing the gate speed limitations that arise due to the off-resonant carrier. Building on this discussion, we then introduce a scheme of micromotion-enabled state-dependent $\sigma^\phi$-forces, which can overcome the limitations on the gate speed due to the off-resonant carrier, provided that the excess micromotion is accurately compensated. 

\subsection{Micromotion effects in the laser-ion interaction}
\label{sec:mumotion_sidebands}

Let us consider a collection of $N$ trapped ions   subjected  to laser beams tuned close to the resonance of a particular transition of frequency $\omega_0$ between two electronic levels $\ket{{\uparrow}_i},\ket{{\downarrow}_i}$. The dynamics of the internal and motional degrees of freedom of this ion crystal is described by the following Hamiltonian
\begin{widetext}
\beq
\label{eq:H0(t)}
H_0(t)=\sum_i\frac{\omega_0}{2}\sigma_i^z+\sum_{i,\alpha}\left(\frac{\hat{p}_{i,\alpha}^2}{2M}+\frac{1}{2}K_\alpha(t) \delta\hat{r}_{i,\alpha}^2-MF_{i,\alpha}(t)\delta\hat{r}_{i,\alpha}+\sum_j\mathbb{V}^\alpha
_{ij}\delta\hat{r}_{i,\alpha}\delta\hat{r}_{j,\alpha}\right),
\eeq
\end{widetext}
where we have introduced  $\sigma_i^z=\ket{{\uparrow}_i}\bra{{\uparrow}_i}-\ket{{\downarrow}_i}\bra{{\downarrow}_i}$,  applied the harmonic-crystal approximation described in the previous section~\eqref{eq:harmonic_approx}, and neglected an irrelevant $c$-number stemming from the classical energy of the breathing crystal.
The interaction between the laser beams and the ions is described by 
\beq
\label{eq:laser_ion_int_one}
H_{\rm I}(t)=\sum_{i,l}\frac{\Omega_{l}}{2}\ee^{\ii\phi_l}\sigma_i^+\ee^{\ii\left(\boldsymbol{k}_l\cdot \hat{\boldsymbol{r}}_{i}(t)+(\omega_0-\omega_l) t\right)}+{\rm H.c.},
\eeq
where we have introduced the spin raising  $\sigma_i^+=\ket{{\uparrow}_i}\bra{{\downarrow}_i}$ and lowering $\sigma_i^-=\ket{{\downarrow}_i}\bra{{\uparrow}_i}$ operators.
Here, $l$ labels the different laser beams that are described as classical traveling waves with ${\boldsymbol{k}_l},\omega_l,\phi_l$ being the laser wavevector, frequency, and phase, respectively, and $\Omega_l$ is the Rabi frequency of the particular transition. Typically, one either considers a quadrupole-allowed transition between a groundstate level $\ket{{\downarrow}}$ and a metastable excited level $\ket{{\uparrow}}$, or uses a two-photon Raman scheme   to couple a pair of groundstate levels  $\ket{{\downarrow}}, \ket{{\uparrow}}$ via a  excited level through a far-off-resonant dipole transition. In any case, the quadrupole or Raman Rabi frequencies are constrained to $|\Omega_l|\ll\omega_0+\omega_l$ in order to neglect additional counter-rotating terms in Eq.~\eqref{eq:laser_ion_int_one}.

We note that this expression is obtained in the interaction picture of the bare Hamiltonian~\eqref{eq:H0(t)}, namely $H_{\rm I}(t)=U^\dagger_0(t)H_{\rm I}U_0(t)$, where $U_0(t)=\mathcal{T}\{{\rm exp}(-\ii\int_0^t{\rm d}t'H_0(t'))\}$.  Thus,
after substituting  the position operator in Eqs.~\eqref{eq:full_solution_quantum_crystal}-\eqref{eq:driven_motion_crystal} corresponding to such an interaction picture, we find 
\begin{widetext}
\beq
\label{eq:laser_ion_int}
H_{\rm I}(t)=\sum_{i,l,\alpha}\frac{\Omega_{l,i}^{\alpha}}{2}\sigma_i^+\left(1+\ii\sum_{m}\mathcal{M}^\alpha_{i,m}\eta_{l,m}^{\alpha}\left(a_{m,\alpha}^{\dagger}u_{m,\alpha}(t)+a_{m,\alpha}^{\phantom{\dagger}}u^*_{m,\alpha}(t)\right)+\cdots\right)\ee^{\ii\phi_{l,i}(t)}\ee^{\ii(\omega_0-\omega_l) t}+{\rm H.c.},
\eeq
\end{widetext}
where we have performed a Taylor series in  the Lamb-Dicke parameters  $\eta_{l,m}^{\alpha}=\boldsymbol{k}_l\cdot{\bf e}_\alpha/\sqrt{2M\omega_{m,\alpha}}\ll 1$, and introduced the renormalised Rabi frequencies $\Omega_{l,i}^{\alpha}=\Omega_l{\rm exp}\{-\sum_m(\mathcal{M}^\alpha_{i,m}\eta_{l,m}^{\alpha})^2/2\}$, together with laser  phases that get modulated by  the time-dependence of the breathing crystal and the  excess micromotion
\beq
\phi_{l,i}(t)=\phi_l+\boldsymbol{k}_l\cdot(\boldsymbol{r}_i^0(t)+\boldsymbol{r}_i^{\rm ex}(t)).
\eeq
 Note that for a linear chain, the breathing crystal becomes static $\boldsymbol{r}_i^0(t)=z_i^0{\bf e}_z$~\eqref{eq:linear_periodic_crystal}, such that the phase modulation is only caused by the excess micromotion. Additionally, 
the effect of the intrinsic micromotion on the laser-ion interaction is encoded in the particular time-dependence of the mode functions $u_{m,\alpha}(t)$~\eqref{eq:normal_mod_solutions}, which yield  additional periodic modulations in processes   involving the  creation and annihilation of phonons.

Let us note that the mode functions are  written in Eq.~\eqref{eq:normal_mod_solutions} as  $u_{m,\alpha}(t)=\ee^{\ii\omega_{m,\alpha}t}f^\alpha_{\rm in}(t)$, were $f^\alpha_{\rm in}(t)$ is a function with period $2\pi/\Omega_{\rm rf}$ already written as  a Fourier series. The excess micromotion leads to $f_{\rm ex}^l(t)=\ee^{\ii\phi_{l,i}(t)}$, which is also a periodic function with period $2\pi/\Omega_{\rm rf}$, and could  as well be expressed as a Fourier series with all the possible harmonics at the different frequencies $\ell\Omega_{\rm rf}$.  In this sense, the micromotion introduces a comb of equidistant sidebands in the laser-ion interaction~\eqref{eq:laser_ion_int}, the so-called {\it micromotion sidebands}. These can be exploited by choosing an appropriate detuning of the lasers with respect to the atomic transition 
\beq
\label{resonance_mumotion_sidebands}
\omega_l-\omega_0\approx \ell_\star\Omega_{\rm rf}, 
\eeq
where $\ell_\star\in\mathbb{Z}$ is a certain integer. In most trapped-ion experiments, one sets $\ell_\star=0$~\cite{nist_review}, and compensates the excess micromotion in order to minimize its effects~\cite{excess_mumotion}. The main result of this work is to point out that addressing the first micromotion sideband $\ell_\star=1$, while maintaining the compensation of  excess micromotion, can be advantageous for QIP. In this way, one can exploit the effects of intrinsic micromotion in the laser-ion interaction, and find faster and more accurate schemes for entangling quantum logic gates.
 
\subsection{State-dependent dipole forces and off-resonant carriers}
\label{sec:state-dep_forces}

We now discuss how to induce a state-dependent $\sigma^\phi$-force~\cite{ms_gates} on the ions starting from Eq.~\eqref{eq:laser_ion_int}, and thus taking into account the new effects brought forth by micromotion. We  consider a pair of laser beams $l\in\{1,2\}$ with equal wavevectors $\boldsymbol{k}_1=\boldsymbol{k}_2:=\boldsymbol{k}_{\rm L}$. By selecting the direction of these beams along a certain trap axis  $\boldsymbol{k}_{\rm L}||{\bf e}_{\alpha}$, the laser-ion interaction will only couple the qubits to a particular phonon branch.

We also consider equal laser phases    $\phi_1=\phi_2=:\phi$,  and  equal intensities and polarizations leading to $\Omega_1=\Omega_2=:\Omega$. Conversely, the lasers will have opposite detunings with respect to the atomic transition $\delta:=\omega_1-\omega_0=-(\omega_2-\omega_0)$. Due to these choices,  we can simplify the laser-ion interaction~\eqref{eq:laser_ion_int}  considerably by defining common  Lamb-Dicke parameters $\eta_{1,n}^\alpha=\eta_{2,n}^\alpha=:\eta^\alpha_n$,   dressed Rabi frequencies $\Omega_{1,i}^\alpha=\Omega_{2,i}^\alpha=:\Omega^\alpha_{i}$,  and modulated phases $\phi_{1,i}(t)=\phi_{2,i}(t)=:\phi_i(t)$, where 
\beq
\label{eq:phases_mumotion}
\phi_{i}(t)=\phi+ {k}_{\rm L}r_{i,\alpha}^0+k_{\rm L}r_{i,\alpha}^{\rm driv}(t)\!\!\left(\!\! 1+ \!\sum_{\ell\geq 1}\!\frac{(-1)^\ell 2q_{\alpha}^\ell\cos(\ell\Omega_{\rm rf}t)}{4^\ell((\ell-1)!)^2}\!\!\right)\!\!.
\eeq
 By keeping contributions to first order in the Lamb-Dicke parameter, $H_{\rm I}(t)=H_{\rm c}(t)+H_{\rm s}(t)$,  we identify the terms driving the  carrier  transitions
\beq
\label{eq:carrier}
H_{\rm c}(t)=\sum_{i}\Omega_{i}^{\alpha}\sigma_i^+\ee^{\ii\phi_{i}(t)}\cos\delta t+{\rm H.c.},
\eeq
and the spin-phonon couplings
\beq
\label{eq:motional_sidebands}
H_{\rm s}(t)=\sum_{i,m}\ii \frak{F}^{\alpha}_{i,m}x^\alpha_m\left(a_{m,\alpha}^{\dagger}u_{m,\alpha}(t)+{\rm H.c.}\right)\sigma_i^+\ee^{\ii\phi_{i}(t)}\cos\delta t+{\rm H.c.},
\eeq
where  the dipole forces are
$\frak{F}^\alpha_{i,m}=\Omega_{i}^{\alpha}\mathcal{M}^\alpha_{i,m}k_{\rm L}$, and the $n$-th mode groundstate widths  are $x^\alpha_m=1/\sqrt{2M\omega_{m,\alpha}}$.

We now consider the effects of the excess micromotion~\eqref{eq:phases_mumotion} to leading order in the regime~\eqref{eq:constraint}, namely \beq
\phi_{i}(t)\approx\varphi_{i}+\tilde{\beta}_{i}\cos(\Omega_{\rm rf}t),
\eeq
where we have introduced the parameters
\beq
\label{eq:beta}
\varphi_{i}=\phi+k_{\rm L}(r_{i,\alpha}^0+r_{i,\alpha}^{\rm driv}(0)),\hspace{2ex} \tilde{\beta}_{i} =-k_{\rm L}r_{i,\alpha}^{\rm driv}(0)\frac{q_\alpha}{2}.
\eeq
Using the Jacobi-Anger expansion~\cite{bessel}, one finds
\beq
\label{eq:excess_bessel}
\ee^{\ii\phi_{i}(t)}=\ee^{\ii\varphi_{i}}\sum_{\ell\in\mathbb{Z}}J_\ell(\tilde{\beta}_{i})\ee^{\ii(\ell\frac{\pi}{2}+\ell\Omega_{\rm rf}t)},
\eeq
where $J_\ell(x)$ are the $\ell$-th order Bessel functions of the first class. This is the explicit expression for the  Fourier series that was mentioned above Eq.~\eqref{resonance_mumotion_sidebands}, and leads to a clear picture for  the appearance of the micromotion sidebands at frequencies $\ell\Omega_{\rm rf}$. Depending on the particular value of the laser detuning $\delta\approx\ell_\star\Omega_{\rm rf}$, it is possible to address a particular micromotion sideband~\eqref{resonance_mumotion_sidebands}. Moreover, around each of these micromotion sidebands, there is an additional comb of frequencies representing the secular sidebands that occur at multiples of the secular normal-mode frequencies~\eqref{eq:sec_freq_crystal}. By combining a pair of first secular sidebands,  the spin-phonon couplings of Eq.~\eqref{eq:motional_sidebands} yield the desired  state-dependent force. 

\subsubsection{Secular state-dependent dipole forces}
\label{sec:secular_forces}

The usual approach to obtain a state-dependent force relies on addressing the secular sidebands, $\delta\sim\omega_{\alpha}\ll \Omega_{\rm rf}$~\cite{ms_gates,ms_gates_nist,innsbruck_MS,oxford_MS,nist_MS}, such that $\ell_\star=0$ according to our previous notation~\eqref{resonance_mumotion_sidebands}. By imposing the condition to resolve the micromotion sidebands
\beq
\label{eq:resolved_mumotion_sidebands}
|\Omega_{i}^{\alpha}|\ll\Omega_{\rm rf},
\eeq 
and using the expression in Eq.~\eqref{eq:excess_bessel} for the effects of excess micromotion, and Eq.~\eqref{eq:normal_mod_solutions} for the effects of the intrinsic micromotion,  we find that the secular sidebands~\eqref{eq:motional_sidebands} can be expressed as a Hamiltonian with a state-dependent force
\beq
\label{eq:secular_force}
H_{\rm s}(t)\approx\sum_{i,m}\frak{{F}}^{\rm r}_{i,m}x^\alpha_m\frak{s}_ia_{m,\alpha}^{\dagger}\ee^{\ii\omega_{m,\alpha}t}\cos\delta t+{\rm H.c.},
\eeq
 where we have introduced a  dipole-force strength
\beq
\label{eq:force_amplitude_secular}
\frak{{F}}^{\rm r}_{i,m}=\frac{\Omega_{i}^{\alpha}\mathcal{M}^\alpha_{i,m}k_{\rm L}}{(1-\frac{q_\alpha}{2})}\sqrt{J_0^2(\tilde{\beta}_{i})+\left(\frac{q_\alpha}{4}J_1(\tilde{\beta}_{i})\right)^2},
\eeq
 and the following spin operator 
 \beq
 \label{eq:spin_op_secular}
 \frak{s}_i=\frac{1}{\sqrt{J_0^2(\tilde{\beta}_{i})+\frac{q_\alpha^2}{16}J^2_1(\tilde{\beta}_{i})}}\left(J_0(\tilde{\beta}_{i})\tilde{\sigma}_i^{y}+\frac{q_\alpha}{4}J_1(\tilde{\beta}_{i})\tilde{\sigma}_i^{x}\right).
 \eeq
Here, we have   defined the Pauli matrices in a rotated basis with respect to the $z$-axis
\beq
\label{eq:Pauli_rot}
\begin{split}
\tilde{\sigma}_i^{x}&:=\ee^{\ii\frac{\varphi_{i}}{2}\sigma_i^z}(\sigma_i^++\sigma_i^-)\ee^{-\ii\frac{\varphi_{i}}{2}\sigma_i^z},\\
\tilde{\sigma}_i^{y}&:=\ee^{\ii\frac{\varphi_{i}}{2}\sigma_i^z}(\ii\sigma_i^--\ii\sigma_i^+)\ee^{-\ii\frac{\varphi_{i}}{2}\sigma_i^z}.\\
\end{split}
\eeq
Accordingly, the spin operator~\eqref{eq:spin_op_secular}  shares certain algebraic properties with the rotated Pauli matrices in Eq.~\eqref{eq:Pauli_rot}, namely  $ \frak{s}_i= \frak{s}_i^\dagger$,  $ \frak{s}_i^2= \mathbb{I}$, and $[\frak{s}_i,\frak{s}_j]=0$, which  allow us to interpret Eq.~\eqref{eq:secular_force} as a state-dependent force that pushes the vibrational modes in opposite directions depending on the two eigenstates of the spin operator $ \frak{s}_i=\ket{+_{\frak{s}_{i}}}\bra{+_{\frak{s}_{i}}}-\ket{-_{\frak{s}_{i}}}\bra{-_{\frak{s}_{i}}}$. In the limit of vanishing excess micromotion $\beta_{i}=0$~\eqref{eq:beta},   the phase $\varphi_i\approx \phi+k_{\rm L}z_i^0\delta_{\alpha,z}$, and Eq.~\eqref{eq:secular_force} yields the  aforementioned  $\sigma^\phi$-force of the M$\o$lmer-S$\o$rensen (MS) scheme~\cite{ms_gates} used in several experiments~\cite{ms_gates_nist,innsbruck_MS,oxford_MS,nist_MS}, where $\sigma^\phi=\ii\ee^{\ii\phi}\sigma^+-\ii\ee^{-\ii\phi}\sigma^-$.

Let us note that, in addition to the desired state-dependent force~\eqref{eq:secular_force}, one has to consider the  carrier terms~\eqref{eq:carrier}, which in this regime~\eqref{eq:resolved_mumotion_sidebands} can be expressed as 
\beq
\label{eq:off_carrrier}
H_{\rm c}(t)=\sum_{i}\Omega_{i}^{\alpha}J_0(\tilde{\beta}_{i})\tilde{\sigma}_i^{x}\cos\delta t.
\eeq
This residual carrier does not commute with the dipole force~\eqref{eq:secular_force}, since the rotated Pauli matrices share the same $\frak{su}(2)$ algebra as the original Pauli matrices. Therefore, the carrier and the dipole force will interfere and compromise the simple picture of the normal modes being displaced in opposite directions depending on the spin state. To minimise this undesired effect, the residual carrier must be far off-resonant, which can be achieved by limiting the laser intensity such that 
\beq
\label{eq:far_off_carrrier_secular}
|\Omega_{i}^{\alpha}J_0(\tilde{\beta}_{i})|\ll\delta\sim\omega_\alpha,
\eeq 
and $H_{\rm c}(t)\approx 0$ in a rotating-wave approximation. For vanishingly-small micromotion $\tilde{\beta}_i\to0$, this constraint~\eqref{eq:far_off_carrrier_secular} reduces to the standard condition required to work in the resolved-sideband regime $|\Omega_i^\alpha|\ll\omega_\alpha$.  As a consequence, resolving the secular sidebands  limits the intensity of the state-dependent force~\eqref{eq:force_amplitude_secular} that becomes in this regime
\beq
\label{eq:force_amplitude_secular_compensated}
\frak{{F}}^{\rm r}_{i,m}=\Omega_{i}^{\alpha}\mathcal{M}^\alpha_{i,m}k_{\rm L}\frac{1}{(1-\frac{q_\alpha}{2})},  \hspace{2ex} \tilde{\frak{s}}_i\approx\tilde{\sigma}_i^{y}
\eeq
As discussed in the following section, it puts a constraint on the speed of entangling gates based on  $\sigma^\phi$-forces. Hence, it would be desirable to come up with schemes that yield similar state-dependent forces with milder constraints on their strengths. We now argue that this is possible by exploiting the higher micromotion sidebands.

\subsubsection{Micromotion state-dependent dipole forces}
\label{sec:micromotion_forces}

Let us now discuss how to obtain a state-dependent force by addressing the first micromotion sideband, $\delta=\Omega_{\rm rf}+\tilde{\delta}$, where $\tilde{\delta}\sim\omega_\alpha\ll \Omega_{\rm rf}$, such that $\ell_\star=1$ according to our previous notation~\eqref{resonance_mumotion_sidebands}. Following an analogous derivation to the one above, we find the following Hamiltonian with a
  state-dependent force
\beq
\label{eq:mumotion_force}
H_{\rm s}(t)\approx\sum_{i,m}\frak{\tilde{F}}^{\rm r}_{i,m}x^\alpha_m \tilde{\frak{s}}_ia_{m,\alpha}^{\dagger}\ee^{\ii\omega_{m,\alpha}t}\cos\tilde{\delta} t+{\rm H.c.},
\eeq
where we have introduced a  dipole-force strength
\beq
\label{eq:force_amplitude_micromotion}
\frak{\tilde{F}}^{\rm r}_{i,m}=\frac{\Omega_{i}^{\alpha}\mathcal{M}^\alpha_{i,m}k_{\rm L}}{(1-\frac{q_\alpha}{2})}\sqrt{J_1^2(\tilde{\beta}_{i})+\left(\frac{q_\alpha}{4}J_0(\tilde{\beta}_{i})\right)^2},
\eeq
 and the following spin operator 
 \beq
 \label{eq:spin_op}
 \tilde{\frak{s}}_i=\frac{1}{\sqrt{J_1^2(\tilde{\beta}_{i})+\frac{q_\alpha^2}{16}J^2_0(\tilde{\beta}_{i})}}\left(-J_1(\tilde{\beta}_{i})\tilde{\sigma}_i^{x}+\frac{q_\alpha}{4}J_0(\tilde{\beta}_{i})\tilde{\sigma}_i^{y}\right).
 \eeq
In analogy with the secular forces~\eqref{eq:secular_force}, we can interpret Eq.~\eqref{eq:mumotion_force} as a state-dependent force that pushes the vibrational modes in opposite directions depending on the two eigenstates of the spin operator $ \tilde{\frak{s}}_i=\ket{+_{\tilde{\frak{s}}_{i}}}\bra{+_{\tilde{\frak{s}}_{i}}}-\ket{-_{\tilde{\frak{s}}_{i}}}\bra{-_{\tilde{\frak{s}}_{i}}}$.

The additional carrier term in Eq.~\eqref{eq:carrier} can be expressed in this case as 
\beq
\label{eq:off_carrrier}
H_{\rm c}(t)\approx\sum_{i}\left(\Omega_{i}^{\alpha}J_0(\tilde{\beta}_{i})\tilde{\sigma}_i^{x}\cos\Omega_{\rm rf} t-\Omega_{i}^{\alpha}J_1(\tilde{\beta}_{i})\tilde{\sigma}_i^{y}\cos\tilde{\delta} t\right).
\eeq
In principle, this term  can cause a similar interference with the state-dependent force~\eqref{eq:mumotion_force}, since it does not commute with the spin operator in general~\eqref{eq:spin_op}.  However, if the excess micromotion is minimized  to the level 
\beq
\label{eq:compensation_goal}
\tilde{\beta}_{i}\ll\fourth q_\alpha\ll 1,
\eeq
 one gets $J_0(\tilde{\beta_i})\approx 1$ and $J_{i}(\tilde{\beta}_i)\approx\tilde{\beta}_i$, such that the previous condition~\eqref{eq:far_off_carrrier_secular}  to neglect the off-resonant carrier becomes less stringent. We find that the laser intensity will be limited by
\beq
\label{eq:far_off_carrrier_micro}
|\Omega_{i}^{\alpha}|\ll\Omega_{\rm rf}, \hspace{2ex}|\Omega_i^{\alpha}|\tilde{\beta}_i\ll\tilde{\delta}\sim\omega_\alpha,
\eeq 
and can be thus tuned to larger values in comparison to the secular scheme~\eqref{eq:far_off_carrrier_secular}, where $|\Omega_i^\alpha|\ll\omega_\alpha$.
According to this discussion, the advantage of   the micromotion-enabled scheme in minimising the undesired effects brought up by the off-resonant carrier with respect to the standard secular scheme will be larger the  smaller  $\omega_\alpha/\Omega_{\rm rf}$ and $\tilde{\beta}_i$ can be made in the experiment. This will depend on the particular trap architecture, and the excess micromotion compensation capabilities discussed below. Let us finally note that 
the state-dependent force~\eqref{eq:force_amplitude_micromotion} becomes in this regime
\beq
\label{eq:mic_forces_compensated}
\frak{\tilde{F}}^{\rm r}_{i,m}\approx \Omega_{i}^{\alpha}\mathcal{M}^\alpha_{i,m}k_{\rm L}\frac{q_\alpha}{4(1-\frac{q_\alpha}{2})}, \hspace{2ex} \tilde{\frak{s}}_i\approx\tilde{\sigma}_i^{y}.
\eeq
Comparing the strength of the secular~\eqref{eq:force_amplitude_secular_compensated} and micromotion~\eqref{eq:mic_forces_compensated} forces, one can see that to obtain similar strengths one would need to increase the Rabi frequency in the micromotion-scheme, and thus the laser power, by a factor of roughly $4/q_\alpha$. At this point, it is worth noting that we could have tuned the laser frequencies to  a higher micromotion sideband $\delta=\Omega_{\rm rf}+\ell_\star\tilde{\delta}$ with $\ell_\star>1$. By doing this,  the effect of the off-resonant carrier would be further suppressed $|\Omega_{i}^{\alpha}|\ll\ell_\star\Omega_{\rm rf}$. On the other hand, we would need even higher laser powers, increased by a factor of $4^{\ell_\star}((\ell_\star-1)!)^2/(q_\alpha)^{\ell_\star}$, to achieve forces of the same strength. Even if these laser intensities can be achieved in the laboratory, in this regime the intensity fluctuations could become a limiting factor for the gate performance. Accordingly, we will focus on the first micromotion sideband  in the rest of this work.

\subsection{Entanglement via geometric phase gates}
\label{sec:gates}

 We now discuss how to exploit the longitudinal/transverse phonons to mediate a qubit-qubit interaction capable of generating entanglement  in the presence of micromotion. In order to have a simple description, we make use of the Magnus expansion~\cite{magnus_expansion}, which allows us to express the time-evolution operator in the interaction picture as follows
 \beq
\label{eq:magnus}
U_{\rm I}(t)=\mathcal{T}\left\{\ee^{-\ii\int_0^t{\rm d}t'H_{\rm I}(t')}\right\}=\ee^{\mathcal{A}(t)}.
\eeq
Here, the anti-Hermitian operator $\mathcal{A}(t)=-\mathcal{A}^\dagger(t)$ can be expressed as a series of time integrals over  nested commutators
\beq
\label{eq:magnus_exp}
  \mathcal{A}(t)=-\ii\int_0^t\!\!\!{\rm d}t_1H_{\rm I}(t_1)-\frac{1}{2}\int_0^t\!\!\!{\rm d}t_1\!\!\!\int_0^{t_1}\!\!\!{\rm d}t_2[H_{\rm I}(t_1),H_{\rm I}(t_2)]+\cdots,
\eeq
which can be truncated to the desired order of approximation. This will allow us to discuss the  generation of entanglement through a generic Hamiltonian with a state-dependent force $H_{\rm I}(t)=H_{\rm s}(t)$, which  encompasses both Eq.~\eqref{eq:secular_force} and Eq.~\eqref{eq:mumotion_force},  and   allows for an additional pulse shaping on the forces $\frak{F}_{i,m}^{\rm r}\to\frak{F}_{i,m}^{\rm r}(t)$. 

In this ideal situation, the Magnus expansion~\eqref{eq:magnus_exp} becomes exact already at second order, such that    
\beq
\label{eq:magnus_op_generic}
\mathcal{A}(t)=\sum_{i,m}\frak{s}_i\left(\gamma_{i,m}(t)a^{\phantom{\dagger}}_{m,\alpha}-\gamma^*_{i,m}(t)a^\dagger_{m,\alpha}\right)+ \sum_{i,j} g_{ij}(t)\frak{s}_i\frak{s}_j,
\eeq
where we have introduced the following parameters
\beq
\label{eq:alpha}
\gamma_{i,m}(t)=-\ii\int_0^t{\rm d}t_1\frak{F}_{i,m}^{\rm r}(t_1)x^\alpha_m\cos(\delta t_1)\ee^{- \ii\omega_{m,\alpha}t_1},
\eeq
\begin{widetext}
\beq
\label{eq:gij}
g_{ij}(t)=\ii\int_0^t{\rm d}t_1\int_0^{t_1}{\rm d}t_2\sum_m\frak{F}_{i,m}^{\rm r}(t_1)x^\alpha_m\frak{F}_{j,m}^{\rm r}(t_2)x^\alpha_m\cos(\delta t_1)\cos(\delta t_2)\sin\big(\omega_{m,\alpha}(t_1-t_2)\big),
\eeq
\end{widetext}

 Hence, the Magnus expansion operator~\eqref{eq:magnus_op_generic} 
 amounts to a state-dependent displacement of the vibrational modes, followed by an effective spin-spin interaction that is capable of generating the desired entanglement between the trapped-ion qubits. On the contrary, the displacement   will degrade the quality of the quantum logic gate, as it leads to  residual entanglement between the qubits and the phonons,  contributing with a motional error that must be minimised. If $\gamma_{i,m}(t_{\rm g})\approx 0$, the vibrational modes develop a closed trajectory in phase space, returning to the initial state after a particular  gate time $t_{\rm g}$. Along these closed trajectories, the qubits acquire a state-dependent geometric phase that depends on the enclosed phase-space area, and  can be exploited to generate maximally entangled states~\cite{gphgates,ms_gates}.
 
For instance, considering $N=2$  and  an initial state ${\rho_0}=\ket{\psi_0}\bra{\psi_0}\otimes\rho_{\rm th}$, where $\ket{\psi_0}=\ket{{\downarrow_1}{\downarrow_2}}$ is the  state of two qubits after optical pumping, and $\rho_{\rm th}$ is the state of the vibrational modes after laser cooling, the time-evolved state under a secular state-dependent force~\eqref{eq:secular_force} in the limit of $\beta_{i}\ll 1$ becomes ${\rho(t_{\rm g})}=\ket{\psi(t_{\rm g})}\bra{\psi(t_{\rm g})}\otimes\rho_{\rm th}$,
where 
\beq
\label{eq:Bell_state}
\ket{\psi(t_{\rm g})}=\frac{1}{\sqrt{2}}\ket{\downarrow_1\downarrow_2}+\frac{\ii}{\sqrt{2}}\ee^{\ii(\varphi_{1}+\varphi_{2})}\ket{\uparrow_1\uparrow_2}
\eeq
is locally equivalent to a Bell state, and we have assumed that  the laser intensities have values such that  $2g_{12}(t_{\rm g})=-\ii \pi/4$.

 In the following subsections, we will  consider different possibilities of achieving $\gamma_{i,m}(t_{\rm g})\approx 0$, and $g_{ij}(t_{\rm g})=-\ii \pi/8$ for a crystal of two trapped ions. We  start by reviewing the entangling gates that use continuous-wave (CW) state-dependent forces in the secular regime~\eqref{eq:secular_force}. We first describe gate schemes that exploit a single vibrational mode (i.e. bus mode) to mediate the entanglement between the qubits~\cite{ms_gates,ms_gates_nist,innsbruck_MS,oxford_MS,nist_MS}, and discuss the limitations in gate speed arising from the necessity to resolve single vibrational modes. We then consider schemes that address both vibrational modes using a CW secular force, and discuss the limitations on the speed imposed by the minimization of spin-motion entanglement of both bus modes. Finally, we move onto a discussion of pulsed schemes, which can overcome both limitations on the gate speed, but will be limited by the restriction on the Rabi frequencies~\eqref{eq:far_off_carrrier_secular}  to neglect the additional off-resonant carrier. This long discussion will allow us to embark upon the description of entangling gates using the micromotion state-dependent-forces of Sec.~\ref{sec:micromotion_forces}, assuming that Eq.~\eqref{eq:compensation_goal} is fulfilled, and discussing the improvement on the gates that this scheme can lead to.

\subsubsection{Entangling gates with secular forces}
\label{sec:secular}

Let us particularize the Magnus  operator~\eqref{eq:magnus_op_generic} to the secular state-dependent dipole force~\eqref{eq:secular_force}, which we will assume to be composed of a sequence of $N_{\rm p}$ square pulses
\beq
\label{eq:pulsed_forces}
\frak{F}^{\rm r}_{i,m}(t)=\sum_{n_{\rm p}=1}^{N_{\rm p}}\frak{f}_{i,m}^{n_{\rm p}}\left(\theta(t-t_{n_{\rm p}})-\theta(t-(t_{n_{\rm p}}+\tau_{n_{\rm p}}))\right).
\eeq
Here, $\frak{f}_{i,m}^{n_p}$ is the force of the $n_{\rm p}$-th pulse obtained from  Eq.~\eqref{eq:force_amplitude_secular}  by substituting the  Rabi frequency $\Omega^{\alpha}_{i,n_{\rm p}}$ of that particular pulse, and $\theta(x)$ is the Heaviside step function, such that  this pulse acts within a time window  $t\in[t_{n_{\rm p}},t_{n_{\rm p}}+\tau_{n_{\rm p}})$. In  order to use the above generic Magnus expansion,  the additional off-resonant carrier~\eqref{eq:off_carrrier} must be negligible, which requires the laser parameters to lie in the regime~\eqref{eq:far_off_carrrier_secular}. Moreover, we will focus on the regime where the excess micromotion is very-well  compensated, such that Eq.~\eqref{eq:far_off_carrrier_secular} leads to  $|\Omega_{i}|\ll\omega_\alpha$, and  thus  $|\frak{f}^{n_{\rm p}}_{i,m} x_m|\ll (\delta+\omega_m)$ follows from Eq.~\eqref{eq:force_amplitude_secular}. This constraint over the forces allows us to simplify considerably the particular expressions for the parameters~\eqref{eq:alpha} and~\eqref{eq:gij} for single- and multi-pulse gates.

\vspace{1ex}
{\it (i) Single-pulse entangling gates:} Let us consider Eq.~\eqref{eq:pulsed_forces} for a single pulse $n_{\rm p}=N_{\rm p}=1$ of strength $\frak{f}_{i,m}$, between $t_{n_{\rm p}}=0$ and $\tau_{n_{\rm p}}=t_{\rm g}$~\cite{ms_gates,ms_gates_nist,innsbruck_MS,oxford_MS,nist_MS}. By performing the corresponding integrals, one finds the state-dependent displacements
\beq
\label{eq:alpha_single_pulse}
\gamma_{i,m}(t_{\rm g})\approx\frac{\frak{f}_{i,m}x^\alpha_m}{2}\frac{1-\ee^{\ii(\delta-\omega_{m,\alpha})t_{\rm g}}}{\delta-\omega_{m,\alpha}},
\eeq
and the phonon-mediated spin-spin interactions
\beq
\label{eq:g_single_pulse}
g_{12}(t_{\rm g})=\ii\nsum[1.4]_m\frac{\frak{f}_{1,m}x^\alpha_m\frak{f}_{2,m}x^\alpha_m}{2(\omega_{m,\alpha}^2-\delta^2)}\!\left(\!\!\omega_{m,\alpha} t_{\rm g}+\frac{\omega_{m,\alpha}\sin(\delta-\omega_m)t_{\rm g}}{\delta-\omega_{m,\alpha}}\!\!\!\right)\!\!.
\eeq

\vspace{1ex}
 {\it (a) Addressing a single vibrational mode:} Let us start by considering single-pulse gates that resolve a single  bus mode to mediate the interaction, such as the center-of-mass (CoM) mode $\delta\approx \omega_{1,\alpha}$ of either  longitudinal or transverse vibrations. The condition to resolve a single vibrational mode for a $N=2$ ion crystal is 
 \beq
 \label{eq:constraint_resolved_mode}
 |\frak{f}_{i,m} x^\alpha_m|\ll |\omega_{1,\alpha}-\omega_{2,\alpha}|, 
 \eeq
such that $\gamma_{i,2}(t_{\rm g})\approx 0$ for the remaining vibrational mode, which only acts as a spectator mode. Hence, one only needs to set the gate time $t_{\rm g}$ such that $\gamma_{i,1}(t_{\rm g})=0$ in order to minimise the residual spin-motion entanglement of the bus mode. This is accomplished by setting the following relation between  laser detuning and the gate time
\beq
\label{eq:detuning_gate_time}
t_{\rm g}=2\pi \frac{r_1}{|\delta-\omega_{1,\alpha}|},
\eeq
where $r_1\in\mathbb{Z}^+$~\cite{ms_gates}. The phase-space trajectory defined by   $\gamma_{i,1}(t_{\rm g})=0$, and induced by the displacement operator~\eqref{eq:magnus_op_generic}, corresponds to $r_1$ closed circular loops, such that spin and motional degrees of freedom get disentangled at the end of the gate. Conversely, the two spins can get maximally entangled. Using $2g_{12}(t_{\rm g})\approx- \ii t_{\rm g}J_{12}$, one finds that  the time-evolution operator~\eqref{eq:magnus}-\eqref{eq:magnus_op_generic} can be expressed as 
\beq
\label{eq:spin_spin_unitary}
U_{\rm I}(t_{\rm g})=\ee^{-\ii t_{\rm g}J_{12}\frak{s}_1\frak{s}_2},
\eeq
where we have introduced the spin-spin coupling strengths 
\beq
\label{eq:J_ij}
J_{12}=\omega_{\rm R}\Omega_{1}^{\alpha}\Omega_{2}^{\alpha} \frac{J_0(\beta_{1})J_0(\beta_{2})}{(1-\half q_\alpha)^2}\nsum[1.4]_m\frac{\mathcal{M}^\alpha_{1,m}\mathcal{M}^\alpha_{2,m}}{\delta^2-\omega_{m,\alpha}^2},
\eeq
where $\omega_{\rm R}=k_{\rm L}^2/2M$ is the recoil energy. Considering a negligible excess micromotion $\beta_i\ll 1$, the coupling  becomes $J_{12}\approx (\Omega^\alpha\eta_1^\alpha)^2\omega_{1,\alpha}/2(\delta^2-\omega_{1,\alpha}^2)\approx(\Omega^\alpha\eta_1^\alpha)^2/4(\delta-\omega_{1,\alpha})$ to leading order of the Lamb-Dicke parameter. This coincides with the expression in Ref.~\cite{ms_gates} up to a different definition of their Lamb-Dicke parameter that incorporates the normal-mode displacements.

The condition to generate a maximally-entangled state using Eq.~\eqref{eq:spin_spin_unitary} is
$
J_{12} t_{\rm g}=\pi/4
$,
which sets another constraint between  laser detuning and the gate time
\beq
\label{eq:max_entang}
 \frac{(\Omega^\alpha\eta_1^\alpha)^2}{|\delta-\omega_{1,\alpha}|}t_{\rm g}=\pi.
\eeq
Solving the system of algebraic equations~\eqref{eq:detuning_gate_time} and~\eqref{eq:max_entang} fixes the detuning as a function of the number $r_1$ of closed  loops in phase space, and the Rabi frequency of the transition $\Omega^\alpha$. Accordingly, the gate time can be shown to be 
\beq
\label{eq:gate_vs_rabi}
t_{\rm g}=\frac{\pi}{\Omega^\alpha\eta^\alpha_1}\sqrt{2r_1},
\eeq
 such that the stronger the intensity of the laser is, the larger  $\Omega^\alpha$ becomes, and the faster the entangling gate is, e.g. $t_{\rm g}=\sqrt{2}\pi/\Omega^{\alpha}\eta_1^{\alpha}$ for gates based on 1-loop trajectories.  We note that this intensity increase must be accompanied by a corresponding increase in the detuning  of the laser beams 
 \beq
 \label{eq:opt_detuning_CoM}
 \delta=\omega_\alpha+\sqrt{2r_1}\Omega^\alpha\eta^\alpha_1.
 \eeq
  However, such an increase in gate speed cannot be prolonged indefinitely. Let us recall that the condition to resolve a single vibrational mode~\eqref{eq:constraint_resolved_mode} puts a constraint on the laser intensity $\Omega\eta_1^\alpha\ll |\omega_{1,\alpha}-\omega_{2,\alpha}|$, such that the gate speed is limited by 
 \beq
 \label{eq:gate_speed_limit_single_mode}
 t_{\rm g}\gg \frac{\pi}{|\omega_{1,\alpha}-\omega_{2,\alpha}|}.
 \eeq
 This gate-speed limitation is very different for MS gates that use  longitudinal or transverse phonons as the quantum bus to mediate the qubit-qubit entanglement.
 
\begin{figure}
\centering
\includegraphics[width=0.9\columnwidth]{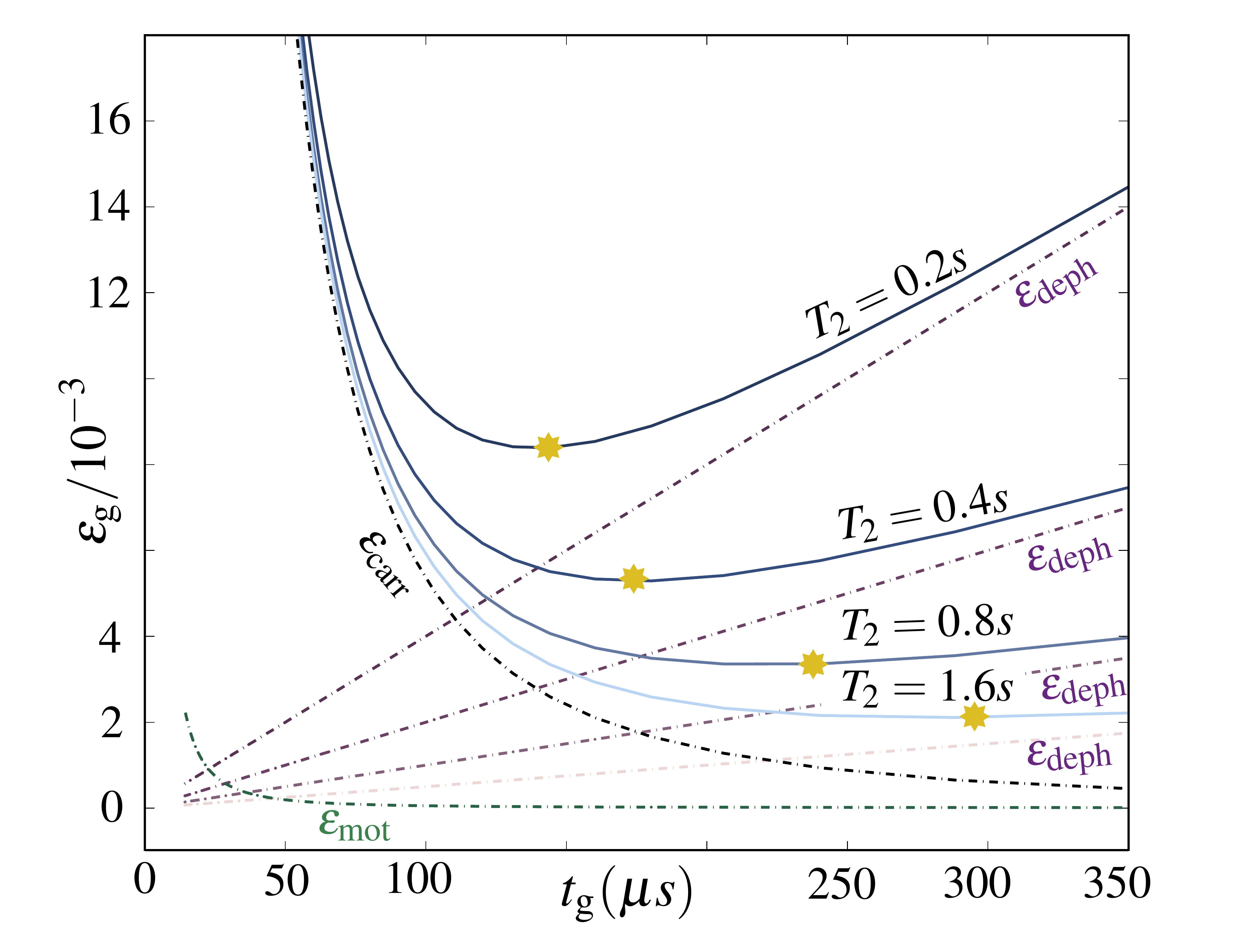}
\caption{ {\bf Single-pulse MS gate with the axial CoM mode:} State infidelity $\epsilon_{\rm g}$  for a MS gate mediated by the CoM longitudinal mode of a  $N=2$  crystal  of  $^{40}{\rm Ca}^+$ ions. We assume an axial trap frequency $\omega_{ z}/2\pi=0.975$MHz, such that the single-ion Lamb-Dicke parameter is $\eta_1^z=0.098$, resolved-sideband  laser-cooling leading to $\bar{n}_z=0.1$ for the CoM mode, and set the number of phase-space loops of the MS gate to $r_1=1$. The blue solid lines correspond to the total state infidelity $\epsilon_{\rm g}$ for  dephasing times $T_2\in\{0.2,0.4,0.8,1.6\}s$, whereas the dotted lines represent the contributions of dephasing, motional, and carrier errors, as indicated in the captions. The yellow stars represent the optimum gate times with respect to the highest-possible gate fidelity for each set of parameters.  To vary the gate speed, we consider increasing the  Rabi frequency within $\Omega^z/2\pi\in[0.02,0.12]$MHz, and setting the corresponding detunings $(\delta-\omega_z)/2\pi\in[2.9,16.6]$kHz according to the equations discussed in the text. }
\label{fig_sp_MS_long}
\end{figure}

 {\it (a.1)} For {\it longitudinal phonons}, the modes fulfill $|\omega_{1,z}-\omega_{2,z}|\sim\omega_z$, such that the gate speed is ultimately limited by the trap period $t_{\rm g}\gg 2\pi/\omega_z$. Let us emphasize, however, that the gate fidelity would decrease  for such fast gates, which sets a lower speed limit in practice.   Maximising the gate speed by increasing the Rabi frequency within the valid parameter regime~\eqref{eq:constraint_resolved_mode}, namely $\Omega^z\ll |\omega_{1,z}-\omega_{2,z}|/\eta_1^z\sim\omega_z/\eta_1^z$, can lead to situations where the contribution of the off-resonant carrier~\eqref{eq:off_carrrier} is not negligible anymore, i.e.  $\Omega^z\ll \omega_z$ in Eq.~\eqref{eq:far_off_carrrier_secular} begins to be violated. Accordingly, if the gate speed increases beyond a certain limit, the off-resonant carrier will increase  the gate  error  and dominate over other sources of noise. 
 
 To quantify this effect, we estimate the state infidelity $\epsilon_{\rm g}=1-\mathcal{F}_{\rm g}$ for the generation of  the desired Bell state~\eqref{eq:Bell_state} as a function of the gate time. We consider three different sources of infidelity $\epsilon_{\rm g}=\epsilon_{\rm carr}+\epsilon_{\rm mot}+\epsilon_{\rm deph}$: the off-resonant carrier~\eqref{eq:off_carrrier} leads to $\epsilon_{\rm carr}\approx \half N(\Omega^z/\delta)^2$~\cite{ms_gates,comment_error}, the additional terms neglected in the Lamb-Dicke expansion~\eqref{eq:laser_ion_int}, including the effect of spectator modes, lead to a motional error $\epsilon_{\rm mot}\approx 0.8\pi N(\delta-\omega_z)(\bar{n}_z+1)/(2\omega_z^2t_{\rm g})+\pi^2N(N-1)(\eta_1^z)^4(1.2\bar{n}_z^2+1.4\bar{n}_z)/8N^2$~\cite{ms_gates}, where we have assumed a thermal state for the longitudinal vibrational modes, such that $\bar{n}_z$ is the mean number of phonons in the thermal CoM mode. Finally, we also consider dephasing during the gate, which can be caused by fluctuating global magnetic fields, which lead to $\epsilon_{\rm deph}\approx 2N^2t_{\rm g}/T_2$, where $T_2$ is the dephasing time of the qubits, as measured by Ramsey interferometry.  In Fig.~\ref{fig_sp_MS_long}, we represent the full error as a function of the gate time for different  dephasing rates, choosing  $^{40}{\rm Ca}^+$ qubits as a representative case~\cite{qip_innsbruck}. This figure demonstrates that for  cold crystals with $\bar{n}_z=0.1$, the motional error of MS gates is negligible in comparison to the errors due to the dephasing and the off-resonant carrier. This would occur also for warmer crystals with the same parameters, provided that $\bar{n}_{z}\leq 5$, after which the motional-error contribution cannot be neglected any longer.  Whereas for slow gates, the  $\epsilon_{\rm deph}$ contribution is dominant, $\epsilon_{\rm carr}$ becomes the leading source of infidelity when the gate becomes  sufficiently fast.  As predicted above, the gate is always slower than the trap period $T_{\rm t}=2\pi/\omega_z=1\mu s$ if one aims for reasonably-high fidelities $\epsilon_{\rm g}<10^{-2}$. In  Sec.~\ref{sec:mic_gates}, we will discuss how it is possible to increase the gate speed further, while maintaining high fidelities,  provided that the intrinsic axial micromotion of the ion crystal can  be exploited to shape the micromotion state-dependent forces~\eqref{eq:mumotion_force} instead of the secular ones~\eqref{eq:secular_force}.

\begin{figure}
\centering
\includegraphics[width=0.9\columnwidth]{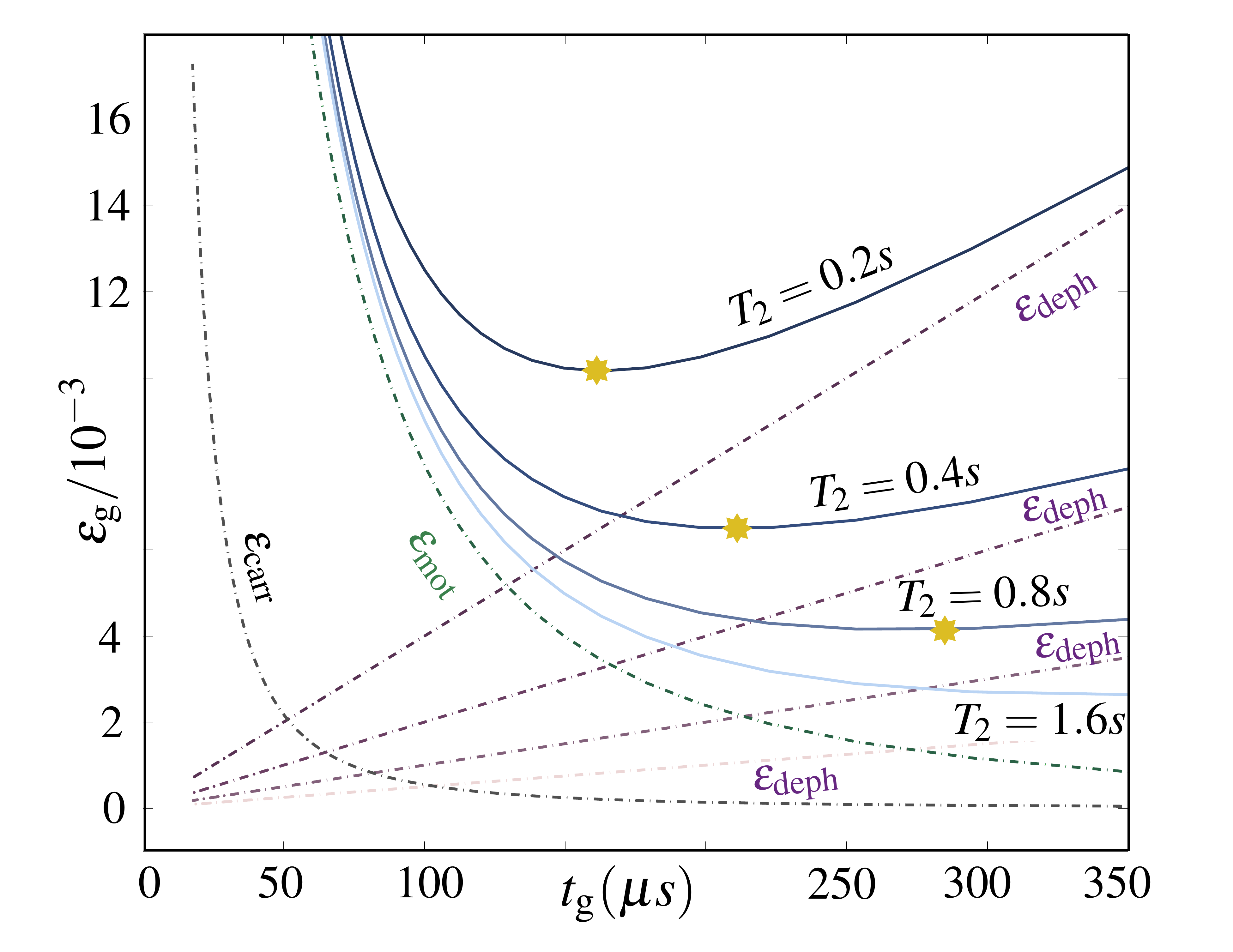}
\caption{ {\bf Single-pulse MS gate with the transverse CoM mode:} State infidelity $\epsilon_{\rm g}$  for a MS gate mediated by the CoM transverse mode of a  $N=2$  crystal  of  $^{40}{\rm Ca}^+$ ions. We assume an axial (radial) trap frequency $\omega_{ z}/2\pi=0.975$MHz, ($\omega_{ x}/2\pi=9.75$MHz) such that the single-ion Lamb-Dicke parameter is $\eta_1^x=0.031$, resolved-sideband  laser-cooling leading to $\bar{n}_x=0.05$ for the CoM mode, and set the number of phase-space loops of the MS gate to $r_1=1$. The blue solid lines correspond to the total state infidelity $\epsilon_{\rm g}$ for  dephasing times $T_2\in\{0.2,0.4,0.8,1.6\}s$, whereas the dotted lines represent the contributions of dephasing, motional, and carrier errors, as indicated in the captions. The yellow stars represent the optimum gate times with respect to the highest-possible gate fidelity for each set of parameters. To vary the gate speed, we consider increasing the  Rabi frequency within $\Omega^x/2\pi\in[0.05,1.29]$MHz, and setting the corresponding detunings $(\delta-\omega_x)/2\pi\in[2.3,56.6]$kHz according to the equations discussed in the text.   }
\label{fig_sp_MS_trans}
\end{figure}
  
 {\it (a.2)} For {\it transverse phonons},   the situation can first appear  favorable, since the trap frequencies are larger, and  one can naively expect that $\delta\sim\omega_x$ can be achieved with  larger detunings, and thus shorter gate times~\eqref{eq:detuning_gate_time}. However, the condition to resolve a single mode~\eqref{eq:constraint_resolved_mode} is more stringent, since  $|\omega_{1,x}-\omega_{2,x}|\sim(\omega_z/\omega_x)^2\omega_x\ll\omega_z$ for the usual regime of linear Paul traps $\omega_z\ll\omega_x$. Therefore, exploiting the available larger detunings to speed up the gate    leads inevitably to a decrease   in  fidelity. We note that the condition to resolve a single mode imposes $\Omega^x\ll |\omega_{1,x}-\omega_{2,x}|/\eta_1^x\ll\omega_z/\eta_1^x\ll\omega_x/\eta_{1,x}$. Hence, even if  the gate speed is  maximised, one would not reach the regime where the off-resonant carrier starts to be problematic since $\Omega^x\ll \omega_x$ is always warranted.   Hence, the error for fast MS gates will be dominated by the contribution to the motional error of the spectator modes.
 
  To quantify this discussion, we estimate again the state infidelity $\epsilon_{\rm g}=1-\mathcal{F}_{\rm g}$ for generating the desired Bell state~\eqref{eq:Bell_state} as a function of the gate time. The carrier and dephasing errors have the same expressions as above, whereas the motional error changes due to the proximity of the spectator modes in frequency space. For $N=2$, we get $\epsilon_{\rm mot}\approx (\Omega^x\eta^x_2)^2(2\bar{n}_x+1)(\delta^2+\omega_{2,x}^2)/(\delta^2-\omega_{2,x}^2)^2$, where we have assumed a thermal state for the transverse vibrational modes with mean phonon number $\bar{n}_x$. In Fig.~\ref{fig_sp_MS_trans}, we represent the full error as a function of the gate time for different  dephasing rates, choosing  $^{40}{\rm Ca}^+$ qubits to compare with the previous longitudinal gate. As announced earlier, this figure shows that the error of slow (fast) gates is dominated by the the dephasing (motional) error. One also observes, that the optimum transverse gates are always slower than the longitudinal ones  in Fig.~\ref{fig_sp_MS_long} and, moreover, achieve smaller fidelities.
 
Let us also note that both of these longitudinal and transverse entangling gates can be generalised to $N$-qubits, and would lead to multi-partite maximally-entangled states locally-equivalent to $\ket{\rm GHZ}_N=(\ket{\downarrow_1\downarrow_2\cdots\downarrow_N}+\ket{\uparrow_1\uparrow_2\cdots\uparrow_N})/\sqrt{2}$, instead of the Bell state~\eqref{eq:Bell_state}. The conditions to generate such states using MS gates based on the longitudinal CoM mode remain the same, since such a bus mode is always separated from higher-frequency modes by the same frequency gap~\cite{normal_modes}. On the contrary, the conditions on MS gates based on the transverse CoM mode lead to even slower gates, since the phonon branch becomes denser, and the different modes approach the CoM frequency as $N$ increases.

\vspace{1ex}
 {\it (b)  Addressing both vibrational modes:} Let us now address how to increase the gate speed  by lifting the constraint~\eqref{eq:constraint_resolved_mode}, such that the state-dependent force does not resolve a single vibrational mode even when $\delta\approx \omega_{1,\alpha}$. For $N=2$ ions,  two conditions are required  to minimize the spin-motion entanglement, namely $\gamma_{i,1}(t_{\rm g})=0$, and $\gamma_{i,2}(t_{\rm g})= 0$. The first one sets the relation in Eq.~\eqref{eq:detuning_gate_time} between the gate time and the detuning, whereas the  yields a commensurability condition 
 \beq
 \label{eq:detuning_opt_2_modes}
 \delta-\omega_{2,\alpha}=r_2(\delta-\omega_{1,\alpha}),
 \eeq
 where $r_2\in\mathbb{Z}$, which already fixes the detuning to 
 \beq
 \label{eq:detuning_two_mdeode_gate}
 \delta=(r_2\omega_{1,\alpha}-\omega_{2,\alpha})/(r_2-1).  
 \eeq
 
{\it (b.1)} For   {\it longitudinal modes}, the  condition~\eqref{eq:detuning_opt_2_modes} cannot be met, as the frequency difference is an irrational number $\omega_{2,z}-\omega_{1,z}=(\sqrt{3}-1)\omega_z$. In any case,  the gate-speed  could not be increased even if  one could close  both trajectories perfectly, as    the  limitation on gate speed
 is given by the condition to neglect the off-resonant carrier (see Fig.~\ref{fig_sp_MS_long}). Equivalently, this would not improve the gate fidelity  of the MS gates too much, as the motional error due to the spectator vibrational mode is already very small for  typical  experimental values (see Fig.~\ref{fig_sp_MS_long}).
 
\begin{figure}
\centering
\includegraphics[width=0.9\columnwidth]{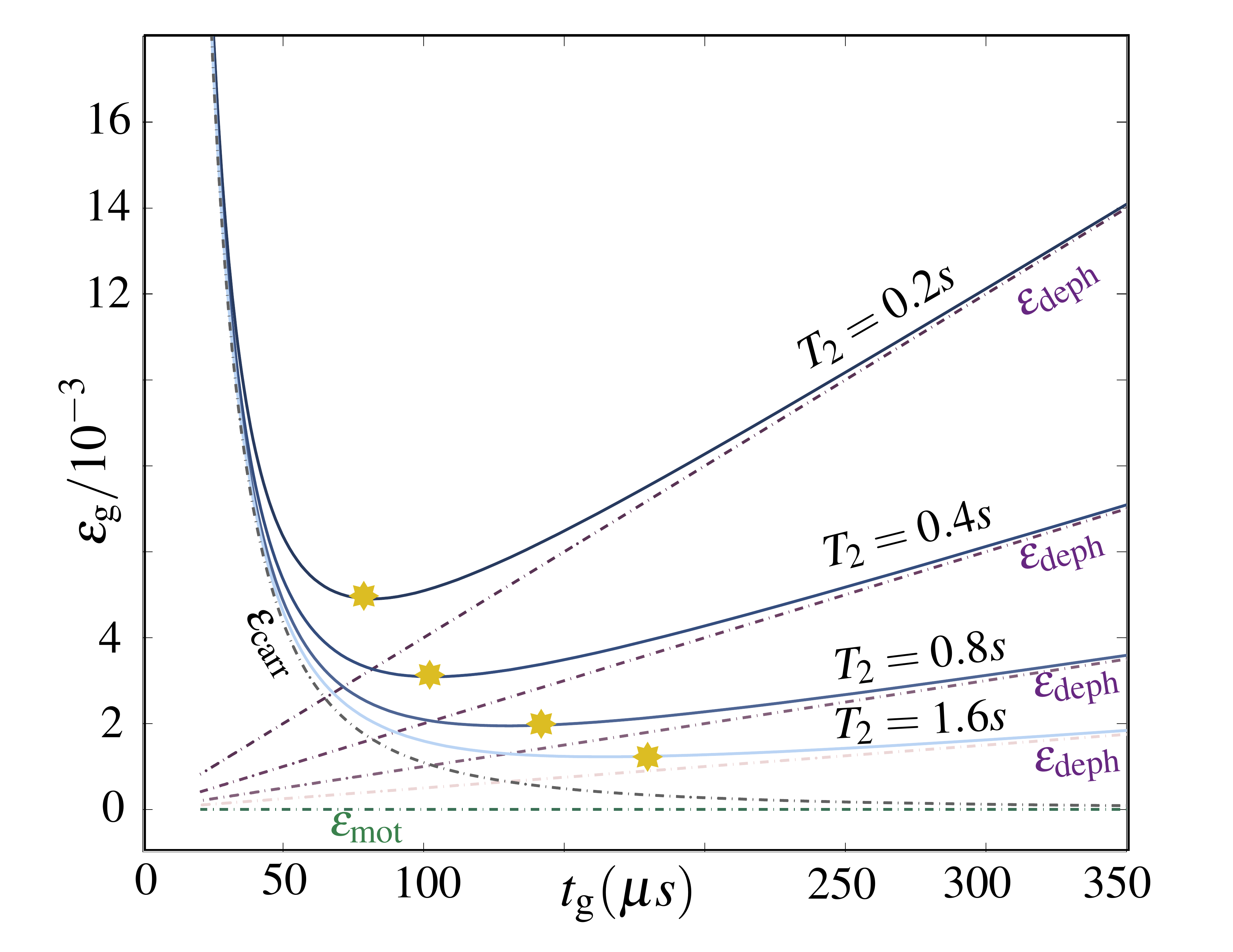}
\caption{ {\bf Single-pulse MS gate with both transverse  modes:} State infidelity $\epsilon_{\rm g}$  for a MS gate mediated by both transverse modes of a  $N=2$  crystal  of  $^{40}{\rm Ca}^+$ ions. We assume a radial trap frequency $\omega_{ x}/2\pi=9.75$MHz leading to the single-ion Lamb-Dicke parameter is $\eta_1^x=0.031$, resolved-sideband  laser-cooling leading to $\bar{n}_x=0.05$ for the CoM mode, and set the number of phase-space loops of the MS gate to $r_1=1$, and $r_2=2$. The blue solid lines correspond to the total state infidelity $\epsilon_{\rm g}$ for  dephasing times $T_2\in\{0.2,0.4,0.8,1.6\}s$, whereas the dotted lines represent the contributions of dephasing, motional, and carrier errors, as indicated in the captions. The yellow stars represent the optimum gate times with respect to the highest  fidelity for each set of parameters. To vary the gate speed, we consider increasing the  axial trap frequency within $\omega_z/2\pi\in[0.2,0.975]$MHz, and setting the corresponding Rabi frequencies $\Omega^x/2\pi\in[0.07,1.58]$MHz and detunings $(\delta-\omega_x)/2\pi\in[2.2,48.9]$kHz according to the equations discussed in the text.   }
\label{fig_sp_MS_trans_2modes}
\end{figure}

{\it (b.2)} For   {\it transverse  modes}, in contrast, the condition~\eqref{eq:detuning_opt_2_modes} can be met and the allowed gate times correspond to trajectories with $r_1$ loops for the center-of-mass mode $\omega_{1,x}$, and $r_1|r_2|$ for the zigzag mode $\omega_{2,x}$ with $r_2\geq 2$ or $r_2\leq-1$.  These  two conditions suffice to fix the gate time to
\beq
\label{eq:gate_time_2_modes}
t_{\rm g}=2\pi \frac{r_1|r_2-1|}{(\omega_{1,x}-\omega_{2,x})}.
\eeq
 The remaining task is to find the required laser intensity such that the state-dependent geometric phase proportional to the enclosed phase-space area fulfills the condition to generate a maximally-entangled state   $ J_{12} t_{\rm g}=\pi/4$. In this case, one has to consider the contribution of both modes to the spin-spin coupling strength~\eqref{eq:J_ij}, which becomes $J_{12}\approx(\Omega^x\eta_1^x)^2(1/4(\delta-\omega_{1,x})-1/4(\delta-\omega_{2,x}))$.  Using the expression for the fixed detuning~\eqref{eq:detuning_two_mdeode_gate}, one finds that the required laser Rabi frequency is 
\beq
\label{eq:opt_Rabi_freq}
\Omega^x\eta^x_1=\frac{(\omega_{1,x}-\omega_{2,x})}{|r_2-1|}\sqrt{\frac{|r_2|}{2r_1|r_2-1|}}.
\eeq
As occurred for the MS gates that use a single vibrational bus mode~\eqref{eq:gate_vs_rabi}, the gate can  become   faster by increasing  the laser Rabi frequency, since the expression
  \beq
\label{eq:gate_vs_rabi_2_modes}
t_{\rm g}=\frac{\pi}{\Omega^x\eta^x_1}\sqrt{\frac{2|r_2|}{r_1|r_2-1|^3}},
\eeq
 yields $t_{\rm g}=2\pi/\Omega^x\eta_1^x$ for the fastest gate with $r_2=2r_1=2$ loops. Let us note that this gate time can also be expressed as 
 \beq
 \label{eq:gate_time_2_modes}
 t_{\rm g}=\frac{2\pi}{\omega_{1,x}-\omega_{2,x}},
 \eeq
which shows  that by exploiting both vibrational modes simultaneously, the  speed  can be increased with respect to the limitation of the previous  transverse MS gates~\eqref{eq:gate_speed_limit_single_mode}.

 We note that the procedure of increasing the gate speed is slightly more involved than that of  single-mode MS gates~\eqref{eq:gate_vs_rabi}, which only required increasing simultaneously the Rabi frequency and the detuning of the laser beams~\eqref{eq:opt_detuning_CoM}. For two-mode MS gates, Eq.~\eqref{eq:opt_Rabi_freq} shows that  in addition one  needs to increase the frequency difference between both vibrational modes, which requires modifying the trap confinement. In particular, we  consider increasing the axial trap frequency $\omega_z$, since  $(\omega_{1,x}-\omega_{2,x})\sim(\omega_z/\omega_x)^2\omega_x$, and this  will increase the gate speed~\eqref{eq:gate_time_2_modes}. The ultimate limit to such an increase in gate speed is caused by the structural instability of the ion chain towards a zig-zag ladder, which occurs for $\omega_z\approx\omega_x$ for $N=2$ ions.  According to Eq.~\eqref{eq:gate_time_2_modes}, this limit corresponds to a gate that could be as fast as the trap period $T_{\rm t}=2\pi/\omega_x$. However, note that the required Rabi frequency in this ultimate limit  would largely violate the condition to neglect the off-resonant carrier~\eqref{eq:far_off_carrrier_secular}, as $\Omega^x\sim(\omega_{1,x}-\omega_{2,x})/\eta_1^x\sim(\omega_z/\omega_x)^2\omega_x/\eta_1^x\sim \omega_x/\eta_1^x\gg\omega_x$. Accordingly, this fast gate would have poor  fidelities. Another effect that would decrease the gate fidelity even further is the increasing importance of non-linear quartic terms in the vibrational Hamiltonian as one approaches the structural instability, which would modify the simple phase-space trajectories of the MS schemes.    
 
 Therefore, at a practical level, the  limit on gate speed for high-fidelity MS gates based on two transverse bus modes
  would be to consider $\omega_z/\omega_x\sim\sqrt{\eta_1^x/10}$, such that  $t_{\rm g}\approx2\pi/(\omega_z/\omega_x)^2\omega_x\gg2\pi/\omega_z\gg2\pi/\omega_x$. Although  this gate is still considerably slower than the trap period, there will be particular ratios $\omega_z/\omega_x$, such that the transverse MS gate may surpass the speed of the longitudinal one. In this sense, by resolving the two vibrational modes, the transverse MS gate can exploit the larger available detunings to achieve higher speeds, while maintaining high fidelities.  
  
  To quantify this discussion, we estimate again the state infidelity $\epsilon_{\rm g}=1-\mathcal{F}_{\rm g}$ for generating the desired Bell state~\eqref{eq:Bell_state} as a function of the gate time. The carrier and dephasing errors have the same expressions as above, whereas the motional error changes once more since  both modes are active buses, and the leading order error will only be caused by the higher-order terms in the Lamb-Dicke expansion~\eqref{eq:laser_ion_int}. For $N=2$, we get $\epsilon_{\rm mot}\approx 2\times\pi^2N(N-1)(\eta_1^x)^4(\bar{n}_x^2+\bar{n}_x)/8N^2$, where we have assumed a thermal state for the transverse vibrational modes with mean phonon number $\bar{n}_x$. In Fig.~\ref{fig_sp_MS_trans_2modes}, we represent the full error as a function of the gate time for different  dephasing rates, choosing  $^{40}{\rm Ca}^+$ qubits to compare with the previous gates. As announced earlier, this figure shows that the error of fast gates is dominated by  the off-resonant carrier error. We note  that   the optimum transverse gates shown in this figure are  faster and more accurate than the longitudinal and transverse gates of Figs.~\ref{fig_sp_MS_long} and~\ref{fig_sp_MS_trans}. Regarding the comparison to the longitudinal-gate performance of Fig.~\ref{fig_sp_MS_long}, we note that  the set of axial trap frequencies used in Fig.~\ref{fig_sp_MS_trans} is always below the axial trap frequency of Fig.~\ref{fig_sp_MS_long} (see the particular values in both captions). Accordingly, the performance and speed shown in  Fig.~\ref{fig_sp_MS_long} sets an upper bound for the comparison of axial and transverse gates, and one concludes that the  the transverse MS gate can indeed achieve higher speeds and fidelities. 
   In  Sec.~\ref{sec:mic_gates}, we will discuss how to increase the gate speed even further, while achieving also higher fidelities,  in traps where the intrinsic radial micromotion can be exploited.

\vspace{1ex}
{\it (ii) Multi-pulse entangling gates:}   In the previous section, we have shown how to increase the speed of  single-pulse MS gates by increasing the laser intensity. Let us now address an alternative strategy to speed up the entangling gates by considering a multi-pulse scheme with $ N_{\rm p}$ pulses~\eqref{eq:pulsed_forces}. In addition to increasing the laser intensity, one can also explore how to distribute it among the different pulses in order to attain higher speeds without compromising the gate fidelities.

To analyse this multi-pulse scheme, we need to find the particular expression for the time-evolution operator in Eqs.~\eqref{eq:magnus} and~\eqref{eq:magnus_op_generic}.  By performing the corresponding integrals in Eqs.~\eqref{eq:alpha} and~\eqref{eq:gij}, we find the following state-dependent displacements and phonon-mediated interaction strengths
\beq
\begin{split}
\gamma_{i,m}(t_{\rm g})&=\Delta \gamma_{i,m}^{N_{\rm p}},\hspace{4ex} \Delta \gamma_{i,m}^{n_{\rm p}}=\sum_{n'_{\rm p}=1}^{n_{\rm p}}\!\gamma_{i,m}^{n'_{\rm p}},
\\
g_{12}(t_{\rm g})&=-\ii \frac{J_{12}^{n_{\rm p}}}{4}t_{\rm g} +\delta g_{12}+\sum_{m\!\phantom{1}}^{\phantom{N_p}}\!\sum_{n_{\rm p}=1}^{N_{\rm p}}\!\!\Delta \gamma_{2,m}^{n_{\rm p}-1}\big(\!\gamma_{1,m}^{n_{\rm p}}\big)^*-{\rm H.c.}.
\end{split}
\eeq
Here, we have introduced the following constants
\begin{widetext}
\beq
\label{eq:parameters}
\begin{split}
\gamma_{i,m}^{n_{\rm p}}&=\frac{\frak{f}^{n_{\rm p}}_{i,m}x^\alpha_m}{2}\left(C^{\tau_{n_{\rm p}}}_{\delta-\omega_{m,\alpha}}\ee^{\ii(\delta-\omega_{m,\alpha})t_{n_{\rm p}}}+C^{\tau_{n_{\rm p}}}_{-\delta-\omega_{m,\alpha}}\ee^{-\ii(\delta+\omega_{m,\alpha})t_{n_{\rm p}}}\right),\hspace{5ex} J_{12}^{n_{\rm p}}=\nsum[1.4]_m\frac{\frak{f}^{n_{\rm p}}_{1,m}x_m^\alpha\frak{f}^{n_{\rm p}}_{2,m}x^\alpha_m\omega_{m,\alpha}}{\delta^2-\omega_{m,\alpha}^2},\\
\delta g_{12}&=\nsum[1.4]_{m}\nsum[1.4]_{n_{\rm p}}\!\frac{\frak{f}^{n_{\rm p}}_{1,m}x^\alpha_m\frak{f}^{n_{\rm p}}_{2,m}x^\alpha_m}{8}\!\left(\!\frac{C^{\tau_{n_{\rm p}}}_{-\delta+\omega_{m,\alpha}}+\left(C^{\tau_{n_{\rm p}}}_{\delta+\omega_{m,\alpha}}-C^{\tau_{n_{\rm p}}}_{2\delta}\right)\ee^{\ii2\delta t_{n_{\rm p}}}}{-\delta+\omega_{m,\alpha}}+\frac{C^{\tau_{n_{\rm p}}}_{\delta+\omega_{m,\alpha}}+\left(C^{\tau_{n_{\rm p}}}_{-\delta+\omega_{m,\alpha}}-C^{\tau_{n_{\rm p}}}_{-2\delta}\right)\ee^{-\ii2\delta t_{n_{\rm p}}}}{\delta+\omega_{m,\alpha}}\!\right),
\end{split}
\eeq
\end{widetext}
and  used the circle function, $C^{\tau}_{\omega}=(1-\ee^{\ii\omega \tau})/\omega$~\cite{pulsed_sz_gates}.
As a consistency check, note  that for a single CW pulse $N_{\rm p}=1, t_{n_{\rm p}}=0$,  the terms $\frak{f}^{n_{\rm p}}_{i,m} x_mC^{\tau_{n_{\rm p}}}_{\omega}$ with $\omega\approx\omega_\alpha$ can be neglected 
by  a rotating-wave approximation for   $|\frak{f}^{n_{\rm p}}_{i,m} x^{\alpha}_m|\ll (\delta+\omega_m)$, which follows from Eq.~\eqref{eq:far_off_carrrier_secular}. Accordingly, one gets the simplified expressions in Eqs.~\eqref{eq:alpha_single_pulse}-\eqref{eq:g_single_pulse}, which were the starting point in the analysis of the previous section.

To illustrate how the CW schemes can be modified to improve the gate speed, we focus on   schemes of equidistant laser pulses of identical widths~\cite{equidistant_schemes,equidistant_schemes_exp}. In this case, one has $\tau_{n_{\rm p}}=\tau:=t_{\rm g}/N_{\rm p}$, and $t_{n_{\rm p}}=\tau (n_{\rm p}-1)$ in Eq.~\eqref{eq:pulsed_forces}. The conditions $\gamma_{i,m}(t_{\rm g})= 0$  yield  a linear system of equations
\beq
\label{eq:sys_equations}
\sum_{n_{\rm p}}{\rm Re}\left\{z_{m,n_{\rm p}}\right\}\Omega^{\alpha}_{i,n_{\rm p}}=0,\hspace{2ex}\sum_{n_{\rm p}}{\rm Im}\left\{z_{m,n_{\rm p}}\right\}\Omega^{\alpha}_{i,n_{\rm p}}=0,
\eeq
where 
 $z_{m,n_{\rm p}}=C^{\tau_{n_{\rm p}}}_{\delta-\omega_{m,\alpha}}\ee^{\ii(\delta-\omega_{m,\alpha})t_{n_{\rm p}}}+C^{\tau_{n_{\rm p}}}_{-\delta-\omega_{m,\alpha}}\ee^{-\ii(\delta+\omega_{m,\alpha})t_{n_{\rm p}}}$, and we denote the Rabi frequencies for each of the pulses as $\Omega^{\alpha}_{i,n_{\rm p}}$.
Therefore, for $N$ ions and thus $N$ normal modes along a particular trap axis, one has a system of $2N$ linear equations, and a non-trivial solution of Eq.~\eqref{eq:sys_equations} can be found if we allow for $N_{\rm p}=2N+1$ different pulses. This solution fixes the relative Rabi frequencies of the pulses $\{\Omega^{\alpha}_{i,n_{\rm p}}/\Omega^{\alpha}_{i,1}\}_{n_{\rm p}=2}^{N_{\rm p}}$. Since we want to study the conditions that allow for   a speed-up with respect to the single-pulse gates  in Eqs.~\eqref{eq:gate_vs_rabi} or~\eqref{eq:gate_time_2_modes}, we shall fix  the detuning to the corresponding optimal value, either Eq.~\eqref{eq:opt_detuning_CoM} or Eq.~\eqref{eq:detuning_two_mdeode_gate} for single/two-mode schemes. Hence, the only remaining equation  comes from the condition to generate a maximally-entangled state $g_{ij}(t_{\rm g})=-\ii \pi/8$. This will suffice to fix $\Omega^{\alpha}_{i,1}$ for a particular gate time, such that we can target pulse sequences that yield faster gates. 

\begin{figure}
\centering
\includegraphics[width=1.\columnwidth]{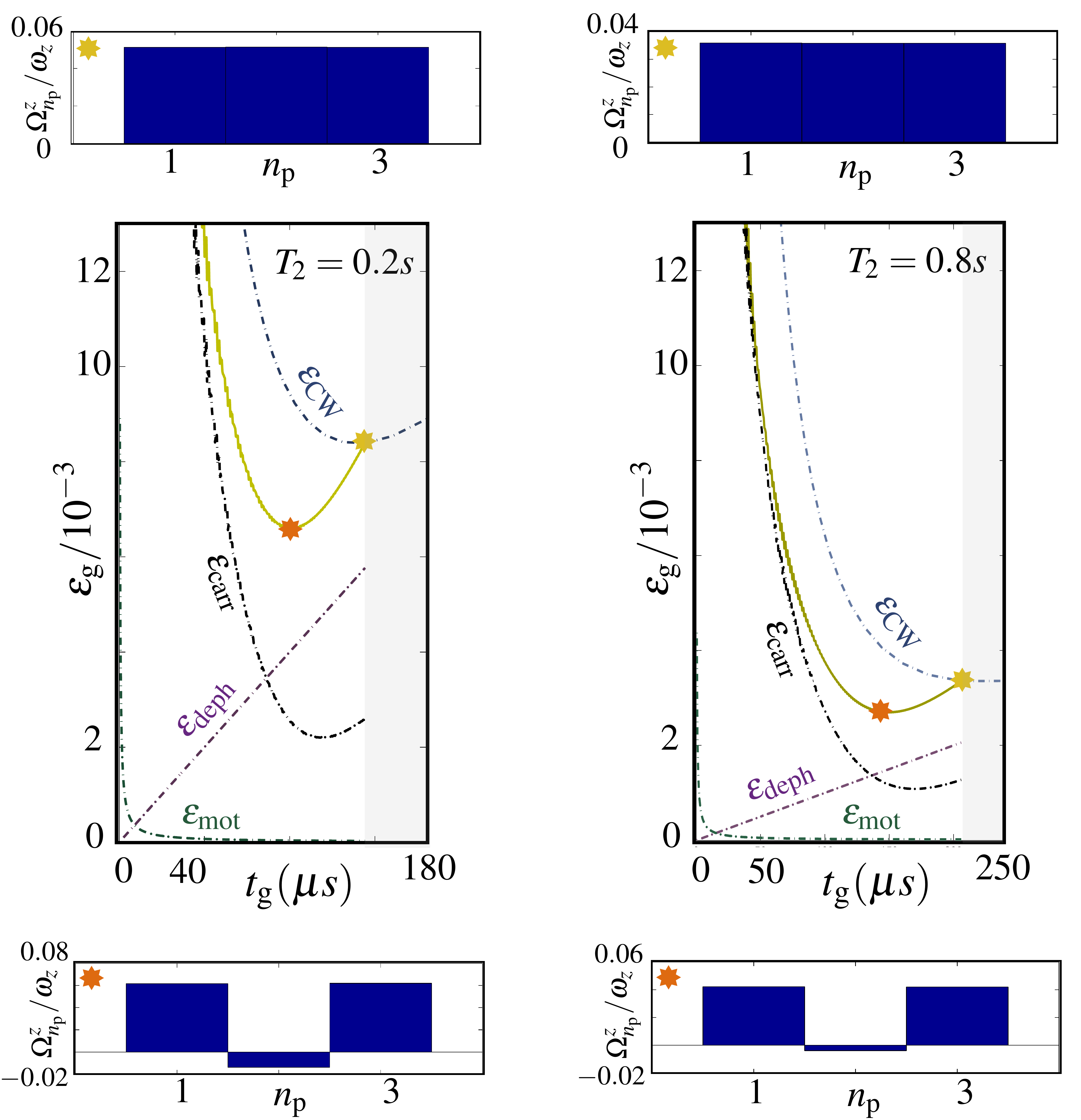}
\caption{ {\bf Multi-pulse MS gate with the axial CoM mode:} (middle panels) State infidelity $\epsilon_{\rm g}$  for a pulsed MS gate mediated by the  longitudinal mode of a  $N=2$  crystal  of  $^{40}{\rm Ca}^+$ ions. We assume an axial trap frequency $\omega_{ z}/2\pi=0.975$MHz, such that the single-ion Lamb-Dicke parameter is $\eta_1^z=0.098$, resolved-sideband  laser-cooling leading to $\bar{n}_z=0.1$ for the CoM mode. The yellow solid lines correspond to the total state infidelity $\epsilon_{\rm g}$ for a $N_{\rm p}=3$-pulsed MS gate under dephasing times $T_2=0.2s$ (left), and $T_2=0.8s$ (right), whereas the dotted lines represent the contributions of dephasing $\epsilon_{\rm deph}$, motional $\epsilon_{\rm mot}$, and carrier $\epsilon_{\rm carr}$ errors, and the gate infidelity of a single-pulse (CW) gate $\epsilon_{\rm CW}$, as indicated in the captions. The yellow stars represent the optimum single-pulse gate times corresponding to   $r_1=1$ phase-space loops. The Rabi frequencies of the pulse train for this regime is given by the upper panels, and coincides with the single-pulse limit.  The orange stars represent the new optimum multi-pulse gate times, obtained by modifying the Rabi frequencies as shown in the lower panels. This different configuration yields  faster and higher-fidelity  gates with respect to the single-pulse cases.}
\label{fig_eq_scheme_axial}
\end{figure}

 {\it (a) Addressing a single vibrational mode:} Let us first address how to increase the  speed of the single-pulse gates based on the longitudinal CoM mode (Fig.~\ref{fig_sp_MS_long}) by exploiting a train of equidistant pulses. For the longitudinal modes, the large frequency gap of the CoM mode with respect to   other vibrational modes allows us to reduce the number of required pulses to $N_{\rm p}=3$. We  follow the above method to find the optimal pulse sequence for a fixed detuning and a certain  gate time. Starting from the gate time of the highest-fidelity MS gates  (see the stars in Fig.~\ref{fig_sp_MS_long}), we  lower the target gate time, and search for pulse sequences that close the CoM phase-space trajectory for a fixed detuning~\eqref{eq:opt_detuning_CoM} that does no longer fulfill Eq.~\eqref{eq:detuning_gate_time}. 
 
\begin{figure}
\centering
\includegraphics[width=1.\columnwidth]{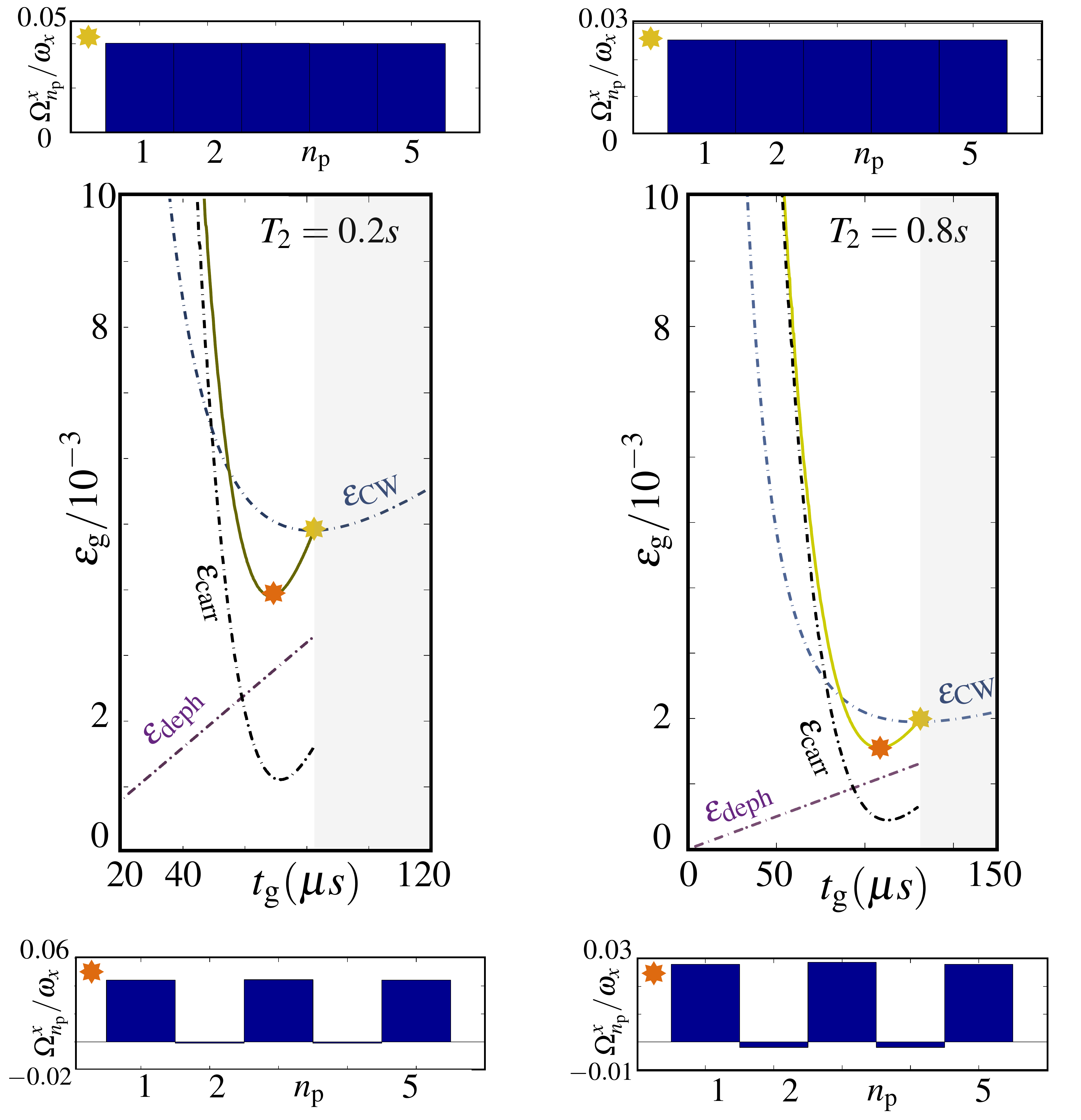}
\caption{ {\bf Multi-pulse MS gate with both transverse  modes:} (middle panels) State infidelity $\epsilon_{\rm g}$  for a pulsed  MS gate mediated by both transverse modes of a  $N=2$  crystal  of  $^{40}{\rm Ca}^+$ ions. We assume a radial trap frequency $\omega_{ x}/2\pi=9.75$MHz leading to the single-ion Lamb-Dicke parameter is $\eta_1^x=0.031$, resolved-sideband  laser-cooling leading to $\bar{n}_x=0.05$ for the CoM mode. The yellow solid lines correspond to the total state infidelity $\epsilon_{\rm g}$ for a $N_{\rm p}=5$-pulsed MS gate under dephasing times $T_2=0.2s$ (left), and $T_2=0.8s$ (right), whereas the dotted lines represent the contributions of dephasing $\epsilon_{\rm deph}$, motional $\epsilon_{\rm mot}$, and carrier $\epsilon_{\rm carr}$ errors, and the gate infidelity of a single-pulse (CW) gate $\epsilon_{\rm CW}$. The yellow stars represent the optimum single-pulse gate times corresponding to   $r_1=1$, $r_2=2$ phase-space loops. The Rabi frequencies of the pulse train for this regime is given by the upper panels, and coincides with the single-pulse limit.  The orange stars represent the new optimum multi-pulse gate times, obtained by modifying the Rabi frequencies as shown in the lower panels. This different configuration yields  faster and higher-fidelity  gates with respect to the single-pulse cases. }
\label{fig_eq_scheme_radial}
\end{figure}

In order to assess quantitatively if the performance of the pulsed MS gate is also optimal, i.e. highest fidelity, we use again the  error model  underlying Fig.~\ref{fig_sp_MS_long}, as discussed in the previous section. However, for the error due to the off-resonant carrier, we consider
  $\epsilon_{\rm carr}\approx \half N\overline{(\Omega^z)^2}/\delta^2$ with $\overline{(\Omega^z)^2}=\sum_{n_{\rm p}}(\Omega^z_{n_{\rm p}})^2/N_{\rm p}$, which  takes into account the distribution of the Rabi frequencies within the pulse train. The results are presented in Fig.~\ref{fig_eq_scheme_axial}, which shows that one can obtain an additional speed-up by using a pulse train with state-dependent forces that alternate their direction. Moreover, the intermediate pulse is very weak, which allows one to reduce the required average Rabi frequency with respect to the single-pulse gates, and leads to a lower gate infidelity. If the multi-pulsed gate speed is increased above this optimum point, the infidelity rises quickly due to the contribution of the off-resonant carrier. This is the main difference with the more-demanding schemes~\cite{fast_gates_kicks,error_repetition_rate,kicked_gates_error,kicks_interferometry,kicks_refocusing} for 
 arbitrary-speed gates that are not based on the resolved-sideband regime~\eqref{eq:laser_ion_int}.
 
\begin{figure*}
\centering
\includegraphics[width=1.6\columnwidth]{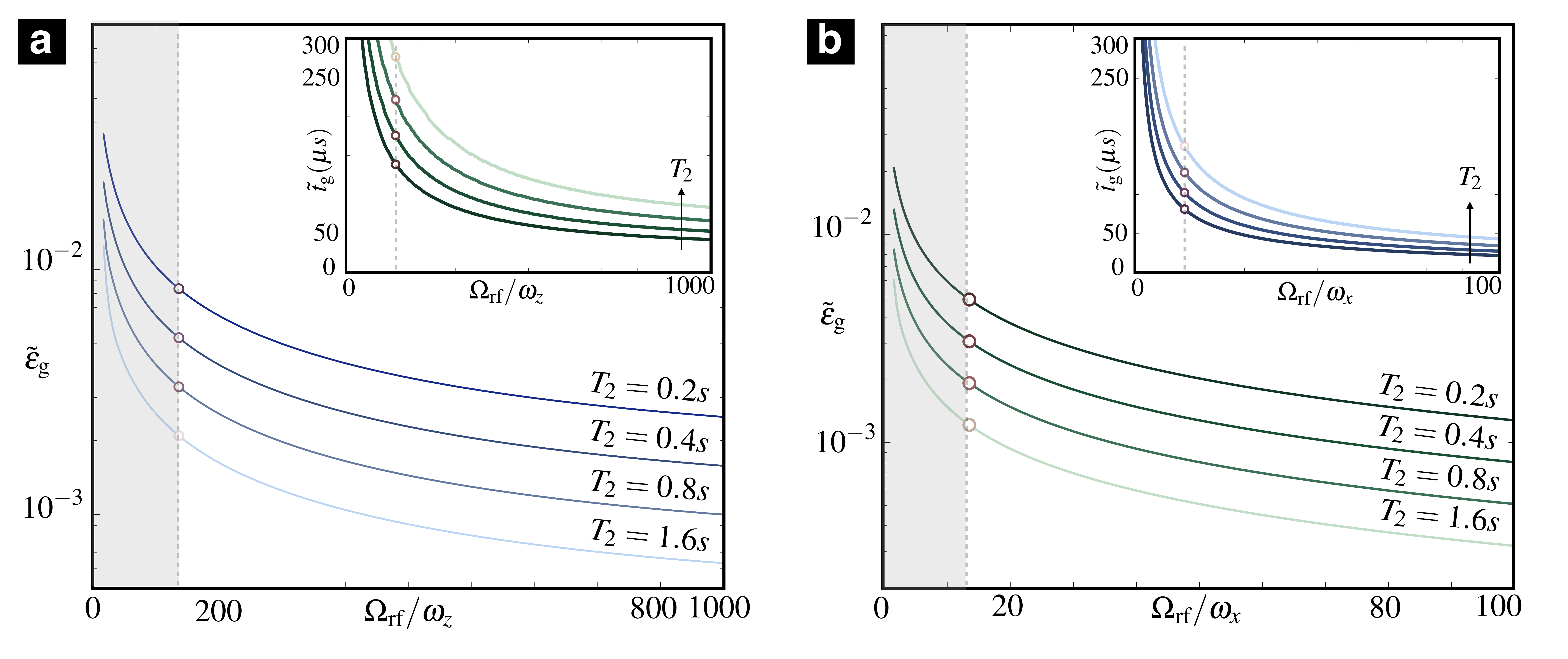}
\caption{ {\bf Micromotion-enabled  improvement of single-pulse  MS gates:} {\bf (a)} (main panel) State infidelity $\tilde{\epsilon}_{\rm g}$  for a single-pulse micromotion MS gate mediated by the longitudinal  modes of a  $N=2$  crystal  of  $^{40}{\rm Ca}^+$ ions with $\omega_{ z}/2\pi=0.975$MHz , $\eta_1^z=0.098$, and $\bar{n}_z=0.1$ for the CoM mode. We consider an axial micromotion parameter $q_z=0.03$, 
and vary the r.f. frequency $\Omega_{\rm rf}$. The dotted line corresponds to $\Omega_{\rm rf}=4\delta/q_z$, and thus  separates the region where the micromotion-enabled MS gates are advantageous (right) and disadvantageous (left shaded region). We represent the corresponding gate times $\tilde{t}_{\rm g}$ in the inset.  The circles in the main panel and inset coincide with the performance of the optimal secular MS gates shown in Fig.~\ref{fig_sp_MS_long}  as yellow stars. Hence, the right region describes a micromotion-enabled improvement in both gate speed and fidelity. {\bf (b)} Same as {\bf (a)} but for a single-pulse  micromotion-enabled MS gate mediated by both transverse modes of a  $N=2$  crystal  of  $^{40}{\rm Ca}^+$ ions with $\omega_{ x}/2\pi=9.75$MHz , $\eta_1^x=0.031$, and $\bar{n}_x=0.05$ for the CoM mode. In this case, we consider   a radial micromotion parameter $q_x=0.3$. Two of the circles in the main panel and inset coincide with the performance of the optimal secular MS gates shown in Fig.~\ref{fig_sp_MS_trans_2modes}  as yellow stars. The micromotion-enabled improvement of these transverse CW gates is qualitatively similar to the longitudinal ones in {\bf (a)}.}
\label{fig_mumotion_CW_gates}
\end{figure*}

{\it (b) Addressing both vibrational modes:} We now study how to increase the gate speed of the single-pulse gates based on both transverse modes (see  Fig.~\ref{fig_sp_MS_trans_2modes}). In this case, the vibrational frequencies are closely spaced, and we need to close all phase-space trajectories using  $N_{\rm p}=2N+1$ pulses. We  follow the above method to find the optimal pulse sequence for a fixed detuning and a certain  gate time. Starting from the gate time of the highest-fidelity MS gates  (see the stars in Fig.~\ref{fig_sp_MS_trans_2modes}), we  lower the target gate time, and search for pulse sequences that close all phase-space trajectories for a fixed detuning~\eqref{eq:detuning_two_mdeode_gate} that does no longer fulfill Eq.~\eqref{eq:gate_time_2_modes}.

In Fig.~\ref{fig_eq_scheme_radial}, we represent the estimated  infidelity  of a $N_{\rm p}=5$-pulse MS gate for $N=2$ ions, as a function of the achieved gate time. In analogy to the axial MS gates in Fig.~\ref{fig_eq_scheme_axial}, we
show that  an additional speed-up can be obtained by  a pulse train with state-dependent forces that alternate their direction (see lower panels). As every-other pulse becomes very weak, we can reduce the required average Rabi frequency with respect to the single-pulse gate, and thus obtain a  higher fidelity. However, increasing the gate speed beyond an optimum point (orange stars) leads to an increase of  the infidelity due to the contribution of the off-resonant carrier.

Although these  results show that the error reduction by moving onto pulsed MS gates is not that large, the improvement in gate speed with respect to the optimal single-pulse gate can be substantial if one only wants to maintain the gate error to the same level. As discussed previously, increasing the  speed even further in both of these pulsed schemes leads to an increase in the infidelity due to the off-resonant carrier. In the following section, we explore the advantage of exploiting the intrinsic micromotion to improve the gate speed even further, while simultaneously maintaining  error rates below a given threshold.

\subsubsection{Entangling gates with micromotion forces}
\label{sec:mic_gates}

 After  this long exposition, we have all the required ingredients to understand how the different MS gate schemes presented above can be improved by exploiting the ion-crystal intrinsic micromotion.  Considering the regime~\eqref{eq:compensation_goal}, one can use directly the previous equations for the secular MS gates discussed in Sec.~\ref{sec:secular}, but taking into account the particular expressions for the micromotion off-resonant carrier~\eqref{eq:off_carrrier} and the micromotion state-dependent forces~\eqref{eq:mic_forces_compensated}.  This simply amounts to substituting in all equations of  Sec.~\ref{sec:secular}: the laser MS detunings by $\delta\to\tilde{\delta}$, the Rabi frequencies  of the secular state-dependent forces by $\Omega_i^\alpha\to \tilde{\Omega}_i^\alpha=\Omega_i^\alpha q_\alpha/4$, and the error due to the off-resonant carrier  by $ \epsilon_{\rm carr}\approx  N\overline{(\Omega^\alpha)^2}/2\delta^2\to\tilde{\epsilon}_{\rm carr}=8 N\overline{(\tilde{\Omega}^\alpha)^2}/q_\alpha^2\Omega_{\rm rf}^2$, where we have further assumed that micromotion compensation fulfils 
 \beq
 \label{eq:beta_compensation}
 \tilde{\beta}_i\ll \frac{q_\alpha\omega_\alpha}{4\Omega_{\rm rf}},
 \eeq
  which is consistent with the experimentally-achieved values that will be discussed  in Sec.~\ref{sec:mumotion_compensation}. This equation gives a practical bound on how small   the excess micromotion must be in order for our analysis to be correct. 

From these substitutions, one observes that the strength of the micromotion state-dependent dipole forces is reduced with respect to the one of the state-dependent secular forces~\eqref{eq:force_amplitude_secular} by a factor of $ q_\alpha/4$. Therefore,   more powerful lasers will be required to achieve the typical speed of  secular MS gates in Figs~\ref{fig_sp_MS_long}-\ref{fig_eq_scheme_radial}. However, provided that such laser sources are available,   the maximum Rabi frequency will not be limited by $|\Omega_i^{\alpha}|\ll\delta$ as occurred for the secular MS scheme~\eqref{eq:far_off_carrrier_secular}, but instead by $|\Omega_i^{\alpha}|\ll\Omega_{\rm rf}$. Hence, exploiting the intrinsic micromotion, one can either maintain the gate speed while increasing the gate fidelity achieved by the secular MS schemes, or vice versa.

 Qualitatively, for the same gate speed, the leading carrier  error for micromotion-enabled MS gates $\tilde{\epsilon}_{\rm carr}$ and   secular MS gates ${\epsilon}_{\rm carr}$ is related by $\tilde{\epsilon}_{\rm carr}={\epsilon}_{\rm carr}(4\delta/q_\alpha\Omega_{\rm rf})^2$. Hence, the carrier error will be reduced provided that 
 \beq
 \label{eq:micromotion_advantage_regime}
\delta < \frac{q_\alpha}{4}\Omega_{\rm rf},
 \eeq
 where $\delta\sim\omega_\alpha$. As announced below Eq.~\eqref{eq:far_off_carrrier_micro}, the advantage  of the scheme will be larger, the  smaller the ratio  $\omega_\alpha/\Omega_{\rm rf}$ can be made in the experiment. The microscopic trap parameter $q_\alpha/4$, which controls the relative amplitude of the intrinsic micromotion and the secular oscillations~\eqref{eq:in_solution}, sets how small is the  ratio $\omega_\alpha/\Omega_{\rm rf}$ required to be for the scheme to be advantageous. From a different perspective, this inequality shows that the coupling to the first micromotion sideband has to be sufficiently big for the scheme to become advantageous.
 
  Conversely, if we want to increase the gate speed but maintain the fidelity of the secular MS gates,  one can show that the gate times $t_{\rm g}$ of single-pulse  secular schemes in Eqs.~\eqref{eq:gate_vs_rabi} or Eq.~\eqref{eq:gate_time_2_modes} are related to the  micromotion-enabled  gate times $\tilde{t}_{\rm g}$ as follows  $\tilde{t}_{\rm g}=t_{\rm g}(4\delta/q_\alpha\Omega_{\rm rf})$. Accordingly, provided that the parameter regime~\eqref{eq:micromotion_advantage_regime} is achieved, there will be a speed-up of the entangling gates. A similar  speed-up will also take place for the multi-pulsed MS gates. 

To be more quantitative, we now study  the total gate infidelity $\tilde{\epsilon}_{\rm g}$  for the micromotion-enabled version of the secular MS schemes of Sec.~\ref{sec:secular}. Therefore, in addition to  the  change in the carrier error already discussed, we also consider the dephasing and motional contributions to the gate infidelity.  We extract the optimal gate time $\tilde{t}_{\rm g}$ that minimizes the gate infidelity $\tilde{\epsilon}_{\rm g}^{\rm min} $, and represent these two quantities as a function of the ratio $\Omega_{\rm rf}/\omega_\alpha$, which determines the region where the micromotion scheme becomes advantageous~\eqref{eq:micromotion_advantage_regime}. 
 
 In Fig.~\ref{fig_mumotion_CW_gates}, we study the micromotion version of the single-pulse secular MS gates mediated by longitudinal (Fig.~\ref{fig_sp_MS_long}) and transverse (Fig.~\ref{fig_sp_MS_trans_2modes}) phonon modes. The circles correspond to r.f. frequencies that fulfill $\Omega_{\rm rf}=4\delta/q_\alpha$, such that the performance of the micromotion-enabled gates coincides with that of the standard secular MS gates. For larger r.f. frequencies (non-shaded regions), the micromotion scheme provides simultaneously lower gate errors (main panel) and lower gate times (inset), both for the MS gates mediated by longitudinal (Fig.~\ref{fig_mumotion_CW_gates} {\bf (a)}) and transverse (Fig.~\ref{fig_mumotion_CW_gates} {\bf (b)}) vibrational bus modes. A similar improvement is found in Fig.~\ref{fig_mumotion_multi-pulsed_gates} for the  micromotion version of the multi-pulse secular MS gates mediated by longitudinal (Fig.~\ref{fig_eq_scheme_axial}) and transverse (Fig.~\ref{fig_eq_scheme_radial}) phonon modes. Let us remark that this micromotion-enabled improvement of multi-pulse MS gates differs from the  results  presented in Ref.~\cite{mumotion_gates_2d}. Here,  C. Shen et {\it al.}  derive sequences for fast entangling gates to mitigate the  adversarial effect of the excess micromotion of planar crystals. In our scheme, we exploit the intrinsic micromotion instead, and turn its effect into a feature that may allow one to improve on both fidelity and speed of phonon-mediated entangling quantum gates.

\begin{figure*}
\centering
\includegraphics[width=1.6\columnwidth]{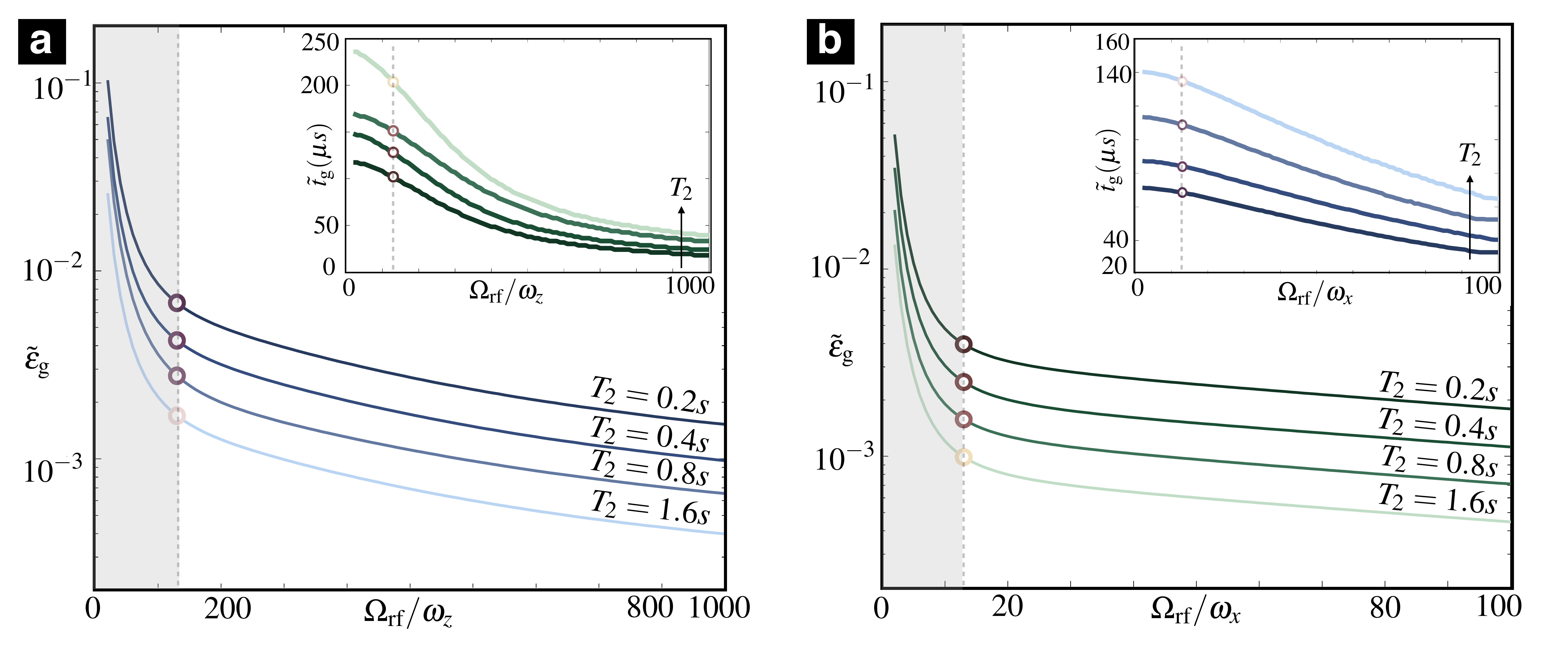}
\caption{ {\bf Micromotion-enabled  improvement of multi-pulse  MS gates:} {\bf (a)} (main panel) State infidelity $\tilde{\epsilon}_{\rm g}$  for a multi-pulse micromotion MS gate mediated by the longitudinal  modes of a  $N=2$  crystal  of  $^{40}{\rm Ca}^+$ ions with $\omega_{ z}/2\pi=0.975$MHz , $\eta_1^z=0.098$, and $\bar{n}_z=0.1$ for the CoM mode. We consider an axial micromotion parameter $q_z=0.03$, 
and vary the r.f. frequency $\Omega_{\rm rf}$. The dotted line corresponds to $\Omega_{\rm rf}=4\delta/q_z$, and thus  separates the region where the micromotion-enabled MS gates are advantageous (right) and disadvantageous (left). We represent the associated gate times $\tilde{t}_{\rm g}$ in the inset.  Two of the circles in the main panel and inset coincide with the performance  of the optimal secular MS gates shown in Fig.~\ref{fig_eq_scheme_axial}  as orange stars. {\bf (b)} Same as {\bf (a)} but for a multi-pulse  micromotion-enabled MS gate mediated by both transverse modes of a  $N=2$  crystal  of  $^{40}{\rm Ca}^+$ ions with $\omega_{ x}/2\pi=9.75$MHz , $\eta_1^x=0.031$, and $\bar{n}_x=0.05$ for the CoM mode. In this case, we consider   a radial micromotion parameter $q_x=0.3$. The circles in the main panel and inset coincide with the performance of the optimal secular MS gates shown in Fig.~\ref{fig_eq_scheme_radial}  as orange stars. 
  }
\label{fig_mumotion_multi-pulsed_gates}
\end{figure*}
 
\section{\bf Experimental considerations}
\label{sec:experimental_cons}

In this section, we discuss the experimental prospects of reaching the required parameter regime that would lead to the micromotion-enabled improvement of the entangling gates described above. 
We start by  discussing in Sec.~\ref{sec:mumotion_compensation} the state-of-the-art excess micromotion compensation, and the possibility of reaching the desired range in Eq.~\eqref{eq:beta_compensation}. In Sec.~\ref{sec:trap_design}, we discuss the difficulty of fulfilling Eq.~\eqref{eq:micromotion_advantage_regime} with current trap designs, and the prospects of satisfying it with realistic trap designs that may become accessible in the future. 
 
\subsection{Compensation of excess micromotion}
\label{sec:mumotion_compensation}

Excess micromotion can give rise to a series of undesired effects~\cite{excess_mumotion}, such as  {\it (i)} a parametric heating  that can increase the secular motion of ion crystals,  limiting the temperatures achieved by laser cooling or even inducing crystal instabilities~\cite{parametric_and_rf_heating}, {\it (ii)} a laser heating for parameters where laser cooling would be expected in the absence of excess micromotion~\cite{parametric_and_rf_heating}, which can also be caused by the intrinsic micromotion~\cite{cooling_intrinsic_mumotion}, and {\it (iii)}  motional shifts of frequency standards (e.g. second-order Doppler shifts)~\cite{excess_mumotion}. Therefore, a great deal of  experimental effort has been devoted over the years to develop methods for a precise estimation and minimization of the excess micromotion. These methods range from  {\it (a)} monitoring the change of the ion equilibrium position as the secular trap frequencies are modified, to {\it (b)} comparing  the  fluorescence intensities of emitted photons when the lasers are tuned either to the bare carrier $\delta\approx 0$ or to the  micromotion carrier $\delta\approx\Omega_{\rm rf}$ (i.e. resolved-sideband regime),  and {\it (c)} monitoring cross correlations of the time delay between the emitted photons and the r.f. signal (i.e unresolved-sideband regime). The precision of method {\it (a)} is limited by the resolution limit of the optics that  measures the ion position, whereas that of {\it (b,c)} depends on limitations and noise on the laser and r.f. sources.

Provided that one of these methods yields an accurate measurement of excess micromotion, one can either apply additional electric fields to compensate the  force of the  spurious d.c. fields~\eqref{eq:forces} due to patch potentials or unevenly coated electrodes, or load the electrodes with reactances to compensate the spurious asymmetries leading to  the oscillating force of the a.c. fields in Eq.~\eqref{eq:forces} (see the discussion in~\cite{excess_mumotion}). A detailed account of the achieved minimization of excess micromotion from different experimental groups can be found in Ref.~\cite{mu_motion_compensation}, which shows that a careful compensation with different methods typically achieves $\beta$-parameters~\eqref{eq:beta} on the order of $\beta_{i}\sim 10^{-3}$. Using tightly-focused dipole beams to probe the ion position can be exploited to achieve even better micromotion compensation~\cite{mumotion_dipole_trap}, so it is reasonable to consider  that the $\beta$-parameter can attain values in the range $\beta_{i}\sim 10^{-4}$-$10^{-3}$. We note that a realistic value for the ideal Paul trap parameters in Eq.~\eqref{eq:a,q} yields $q_\alpha\sim0.2$-$0.3$ for the transverse directions $\alpha=\{x,y\}$,  such that the desired compensation regime in Eq.~\eqref{eq:beta_compensation} can be achieved with state-of-the-art trapped-ion technology. For the axial direction, considering short segmented linear traps, one may achieve ratios of $q_z/q_x \approx 10^{-3}$~\cite{axial_segmented}. Considering the performance of the axial micromotion-enabled entangling gates of Figs.~\ref{fig_mumotion_CW_gates}{\bf (a)} and~\ref{fig_mumotion_multi-pulsed_gates} {\bf (a)} for $q_z=0.03$, the  smaller values of $q_z$ for these segmented traps would require a much higher ratio of $\Omega_{\rm rf}/\omega_{\rm z}$, as well as a much higher laser power to achieve similar gate speeds. Accordingly, finding experimental trap designs that meet the requirements for a micromotion-enabled improvement  based on axial modes seems very challenging, and this motivates us to consider the radial micromotion gates below.

\subsection{Discussion of current and future trap designs }
\label{sec:trap_design}

The suggested scheme requires a large ratio of the drive frequency to the
secular motional frequency in the radial direction $\Omega_{\rm rf} /
\omega_{x}$. Since the confinement properties of ion traps can be accurately  described by 
Mathieu equations that are independent on the actual trap geometry~\cite{Trap_book},  a study based on a geometry that is suitable for usual trapping parameters will
also suffice to explore the possibility of reaching  the required parameters for a micromotion-enabled improvement of the entangling gate. Current
traps for quantum information processing operate usually in the regime
of $\Omega_{\rm rf} / \omega_{x} \approx 10-20$~\cite{sqip_trap,
	trap_resonator}.  Experimentally, multi-qubit gate operations with
$\Omega_{\rm rf} / \omega_{x} = 46$ have already been demonstrated using
$^{40}$Ca$^+$~\cite{alonso_priv, transport_gates}. This ratio, together with the rest of the parameters used in 
Fig.~\ref{fig_mumotion_CW_gates}{\bf (b)},  would already yield a benefit from the  micromotion-enabled entangling  gates. To be more precise, assuming a decoherence time of $T_2=0.8$s and  the error model described above, the single-pulse MS gate based on secular radial forces would reach $\epsilon_{\rm g}=2\cdot 10^{-3}$ in a time  $t_{\rm g}=129\mu$s, whereas the one based on  micromotion radial forces could attain $\epsilon_{\rm g}=8\cdot 10^{-4}$ in a time $t_{\rm g}=57\mu$s. Let us note that to gain full advantage of the protocol, one would
need even higher ratios of $\Omega_{\rm rf} /
\omega_{x}$, which have not been achieved yet in experiments.

While there is no fundamental reason that will prohibit reaching even
higher drive frequencies, one needs to take practical considerations
into account. The dissipated power inside the trap will increase since
the amplitude of the r.f. drive voltage needs to be increased, leading
thus to a higher power dissipation in the trap itself, and also in the
electrical connections to the trap~\cite{trap_resonator}. Managing the
increased heat load will require complex thermal management
techniques, especially in the context of cryogenic systems.  In this
context, a smaller trap and connection capacitance is beneficial as it
will facilitate the design of the required circuitry to generate the
radio frequency trapping fields~\cite{trap_resonator}.

\subsection{Technical noise sources}
Estimating the error budget accounting for additional technical limitations can, for the proposed gate scheme,  be
performed analogously to other high-fidelity entangling gate operations, as detailed for instance in
Ref.~\cite{oxford_MS}. Regarding the differences  for the micromotion-enabled gates,  
let us note that, in case that the experimentally available laser power  is  limited,  the gate
duration would be increased by a factor of $1/\sqrt{q_\alpha}$, which follows from the different scaling of the dipole forces in Eqs.~\eqref{eq:force_amplitude_secular_compensated} and~\eqref{eq:mic_forces_compensated}. In general, this would make the gate more susceptible to
dephasing noise. Accordingly, if laser power is the limiting factor, one should consider  continuous~\cite{continuous_dd, continuous_dd_ions} or pulsed~\cite{DD_MS}  dynamical decoupling techniques to combat this noise.

Another technical aspect that would differ from entangling gates that do not make explicit use of micromotion  is the generation of the
bichromatic light fields. In the presented gate, the frequency of the
beat note must be on the order of $2 \Omega_{\rm rf}\approx 2\pi\, 100\,$MHz,
whereas for the standard gate the modulation frequency is on the order
of $\omega_\alpha/ 2\pi\approx 1$-$10\,$MHz. The modulation is usually generated
using acousto-optical modulators, which are available with a bandwidth
of $100\,$MHz, at the cost of a reduced diffraction efficiency. This
larger detuning from the carrier transition in our scheme brings the additional
advantage that incoherent excitation of the qubit due to residual
laser intensity at the carrier transition is reduced
considerably. This incoherent excitation poses a major problem for
qubits driven by narrow linewidth diode laser
systems~\cite{OzeriLaser}.

\section{\bf Conclusions and Outlook}
\label{sec:conclusions}

In this work, we have developed a set of theoretical tools to analyze the effects of excess and intrinsic micromotion in the schemes for  high-fidelity quantum logic gates with  trapped-ion qubits. We have shown that, in situations where the excess micromotion is compensated to a high degree, it is possible to exploit the  intrinsic micromotion to improve on both the   speed and  fidelity of current schemes for entangling gates. We have derived a set of conditions that identify the parameter regime where such an improvement can occur, and discussed the possible challenges of reaching this regime considering realistic experimental conditions.

Aside from the particular gate scheme, we have presented for the first time a detailed quantum-mechanical treatment of intrinsic and excess micromotion in arbitrarily-large chains of trapped ions. This has allowed us to develop a generic theory for the laser-ion interaction in the presence of micromotion, which might be useful for  future trapped-ion studies in completely different contexts.  

{\it Acknowledgements.--} We thank J. Home, J. Alonso, and T. Mehlst\"{a}ubler for useful conversations, and H. Landa and A. Lemmer for their  comments and the careful reading of the manuscript. 

The research is based upon work supported by the Office of the Director of National Intelligence (ODNI), Intelligence Advanced Research Projects Activity (IARPA), via the U.S. Army Research Office Grant No. W911NF-16-1-0070. The views and conclusions contained herein are those of the authors and should not be interpreted as necessarily representing the official policies or endorsements, either expressed or implied, of the ODNI, IARPA, or the U.S. Government. The U.S. Government is authorized to reproduce and distribute reprints for Governmental purposes notwithstanding any copyright annotation thereon. Any opinions, findings, and conclusions or recommendations expressed in this material are those of the author(s) and do not necessarily reflect the view of the U.S. Army Research Office. 

We also acknowledge support by U.S. A.R.O. through Grant No. W911NF-14-1-010.  A. B. acknowledges support from Spanish MINECO Project FIS2015-70856-P, and CAM regional research consortium QUITEMAD+. P. S., T. M. and R. B.  acknowledge support  from the Austrian Science Fund (FWF), through the SFB FoQus (FWF Project No. F4002-N16) and the Institut f\"{u}r Quanteninformation GmbH.



\begin{references}
\bibitem{comment_error}
This particular expression for the contribution of the off-resonant carrier to the error budget is an upper bound derived in Ref.~\cite{ms_gates}. We note that even if the carrier error is smaller than this value, the general trend discussed in this work will remain, i.e. it will become the leading  error for sufficiently fast gates.
\end{references}
\end{document}